\documentclass[useAMS,usenatbib]{mn2e}

\usepackage{verbatim} 
\usepackage{natbib} 
\usepackage{amsmath} 
\usepackage{amsbsy}
\usepackage{amssymb}
\usepackage{mathrsfs} 
\usepackage{lscape} 
\usepackage{graphicx}
\usepackage{epstopdf}
\usepackage{deluxetable} 
\usepackage{fixltx2e} 

\title[AGN outflow physics]{The physics of galactic winds driven by active galactic nuclei}

\author[Faucher-Gigu\`ere \& Quataert]{Claude-Andr\'e Faucher-Gigu\`ere\thanks{Miller Fellow;
cgiguere@berkeley.edu}, Eliot Quataert \vspace*{6pt}  \\ 
Department of Astronomy and Theoretical Astrophysics Center, University of California, Berkeley, CA 94720-3411, USA}

\begin{document}
\maketitle


\begin{abstract}
Active galactic nuclei (AGN) drive fast winds in the interstellar medium of their host galaxies. 
It is commonly assumed that the high ambient densities and intense radiation fields in galactic nuclei imply short cooling times, thus making the outflows momentum-conserving.  
We show that cooling of high-velocity, shocked winds in AGN is in fact inefficient in a wide range of circumstances, including conditions relevant to ultra-luminous infrared galaxies (ULIRGs), resulting in energy-conserving outflows. We further show that fast energy-conserving outflows can tolerate a large amount of mixing with cooler gas before radiative losses become important. 
For winds with initial velocity $v_{\rm in} \gtrsim 10,000$ km s$^{-1}$, as observed in ultra-violet and X-ray absorption, the shocked wind develops a two-temperature structure.
While most of the thermal pressure support is provided by the protons, the cooling processes operate directly only on the electrons. 
This significantly slows down inverse Compton cooling, while free free cooling is negligible. 
Slower winds with $v_{\rm in}\sim 1,000$ km s$^{-1}$, such as may be driven by radiation pressure on dust, can also experience energy-conserving phases but under more restrictive conditions. 
During the energy-conserving phase, the momentum flux of an outflow is boosted by a factor $\sim v_{\rm in}/ 2 v_{\rm s}$ by work done by the hot post-shock gas, where $v_{\rm s}$ is the velocity of the swept-up material. Energy-conserving outflows driven by fast AGN winds ($v_{\rm in} \sim 0.1c$) may therefore explain the momentum fluxes $\dot{P} \gg L_{\rm AGN}/c$ of galaxy-scale outflows recently measured in luminous quasars and ULIRGs. 
Shocked wind bubbles expanding normal to galactic disks may also explain the large-scale bipolar structures observed in some systems, including around the Galactic Center, and can produce significant radio, X-ray, and $\gamma$-ray emission. The analytic solutions presented here will inform implementations of AGN feedback in numerical simulations, which typically do not include all the important physics.
\end{abstract}


\section{Introduction}
Recent observations show compelling evidence of galaxy-scale outflows driven by active galactic nuclei (AGN). 
These include winds of ionized, neutral, and molecular gas in ultra-luminous infrared galaxies \citep[ULIRGs;][]{2010A&A...518L..41F, 2010A&A...518L.155F, 2011ApJ...729L..27R, 2011ApJ...733L..16S} and quasars \citep[][]{2009ApJ...690..953F, 2009ApJ...706..525M, 2010ApJ...709..611D, 2010ApJ...713...25B, 2010MNRAS.402.2211A, 2011arXiv1108.2392V}. Possibly related fast outflows have also been detected in post-starburst galaxies \citep[][]{2007ApJ...663L..77T, 2011Sci...334..952T} and in Lyman break galaxies \citep[][]{2011ApJ...733...31H}. While quasar feedback has long been postulated to play important roles in galaxy formation, including establishing the $M_{\rm BH}-\sigma$ relation \citep[e.g.,][]{1998A&A...331L...1S, 2003ApJ...595..614W, 2005Natur.433..604D} and truncating star formation in massive galaxies \citep[][]{2005ApJ...620L..79S, 2008ApJS..175..390H}, it is only recently that observations have directly revealed the impact of AGN outflows on galaxy scales. 
Furthermore, the data set on such outflows is poised to explode in the near future as the \emph{Herschel Space Observatory}\footnote{http://herschel.esac.esa.int} continues its mission and the \emph{Atacama Large Millimeter Array}\footnote{http://science.nrao.edu/facilities/alma} (ALMA) becomes operational. 
It is thus important to develop a theoretical understanding of the dynamics of AGN-driven winds.

The most basic dynamical question regarding AGN winds is whether they conserve energy or momentum. 
Although energy is conserved globally, the wind can in principle radiate away the thermal energy generated when it shocks with the surrounding interstellar medium (ISM). On the other hand, momentum cannot be radiated away. 
In the limit in which the shocked wind energy is rapidly radiated away, we term the outflow ``momentum-conserving.'' If radiative losses are negligible, we say that the outflow is ``energy-conserving.'' 
Real outflows may of course be somewhat intermediate between these two limits. 
Nevertheless, the degree to which AGN outflows conserve energy has very significant implications. 
In particular, the momentum flux of the material swept up in energy-conserving outflows increases with time owing to work done by hot shocked gas. 
An analogous phenomenon operates in supernova remnants (SNRs), in which the momentum of the remnant at the end of the Sedov-Taylor (energy-conserving) phase can exceed the momentum of the explosion ejecta by a factor as high as $\sim 50$ \citep[e.g.,][]{1988ApJ...334..252C}. 
Such momentum boosts have a large effect on the efficiency of stellar feedback in galaxies \citep[e.g.,][]{2011ApJ...731...41O, 2012MNRAS.tmp.2654H}. 

Both observations and theoretical modeling suggest that galaxy-scale AGN outflows have momentum fluxes well in excess of the momentum flux output radiatively by the central black hole (BH), $\dot{P}_{\rm rad} = L_{\rm AGN}/c$. 
Observationally, the ``momentum boost'' $\dot{P}/\dot{P}_{\rm rad}$ ranges from $\sim2$ to $\sim 30$ in local ULIRGs dominated by AGN \citep{2011ApJ...729L..27R, 2011ApJ...733L..16S} and in luminous quasars for which the outflow properties have been measured accurately \citep[][]{2009ApJ...706..525M, 2010ApJ...709..611D, 2010ApJ...713...25B, 2011arXiv1108.0413F}. 
The existing measurements are summarized in greater detail in \S \ref{momentum boost sec}. 
Although the uncertainties are large, collectively these measurements indicate that AGN-driven, galaxy-scale outflows may commonly have momentum fluxes $\sim 10 L_{\rm AGN}/c$. 
On the theoretical side, simulations that model AGN feedback by depositing radial momentum on scales $\sim 100$ pc have found that comparably high momentum fluxes are needed to reproduce the normalization of the $M_{\rm BH}-\sigma$ relation \citep[][]{2011MNRAS.412.1341D, 2011MNRAS.tmp.2150D}.  
If AGN outflows are energy-conserving, it is possible that they are radiatively launched with momentum flux $\leq L_{\rm AGN}/c$ in the nucleus, then boosted to $\sim 10 L_{\rm AGN}/c$ on galaxy scales in a Sedov-Taylor-like phase. 

The original work of \cite{1998A&A...331L...1S} on quasar feedback implicitly assumed that the outflow was energy-conserving. 
Subsequently, \cite{2003ApJ...596L..27K} argued that inverse Compton cooling by the quasar radiation field should efficiently cool the shocked wind, making the outflow momentum-conserving out to $\sim1$ kpc, but energy-conserving past this radius \citep[][]{2011MNRAS.415L...6K, 2012arXiv1201.0866Z}. 
\cite{2010ApJ...725..556S} argued that energy-conserving outflows are not possible in galaxy bulges, although the cooling times in their numerical models appear too short in comparison with the analytic theory developed herein. 
In this work, we focus primarily on fast ($\gtrsim10,000$ km s$^{-1}$) nuclear winds and show that cooling is less important than previously appreciated. 

Such winds may be launched from accretion disks around central black holes and give rise to ultra-violet (UV) broad absorption lines \cite[BALs;][]{1995ApJ...451..498M}. 
Observed BALs have typical maximum velocity offsets $\sim10,000-30,000$ km s$^{-1}$, and even higher velocity absorption has been detected in some cases \citep[e.g.,][]{1981ARA&A..19...41W, turnshek88, 2009ApJ...692..758G}. However, we will refer to fast nuclear winds generically in this paper since the winds need not be directly connected to an accretion disk (for example, if driven by radiation pressure on a larger scale). 
Furthermore, fast winds need not manifest themselves in the form of classical BALs. 
X-ray observations also find evidence for AGN winds with velocities $\sim0.1c$, albeit with lower significance and with more ambiguous interpretation \citep[e.g.,][]{2002ApJ...579..169C, 2003MNRAS.346.1025P, 2010A&A...521A..57T}.  
We also briefly consider slower winds ($\sim 1,000$ km s$^{-1}$), as may be accelerated by radiation pressure on dust \citep[e.g.,][]{2012arXiv1204.0063R}. These winds too can experience significant energy-conserving phases, but under more restrictive conditions. 

The cooling of the shocked wind in AGN involves important conceptual issues which have not been previously discussed in this context. 
Furthermore, key physical processes (in particular, inverse Compton cooling and two-temperature plasma effects) are not included in many existing numerical simulations of AGN feedback. 
Simulations of isolated galaxies or galaxy mergers, and most acutely cosmological simulations, also typically do not resolve all the relevant scales, including the wind shock. 
Thus, we focus here on exploring the basic physics of galaxy-scale outflows analytically. 
This work will inform the implementation of AGN winds in future numerical simulations, which will be best suited to address potential three-dimensional effects in more realistic settings.

The plan of this paper is as follows. In \S \ref{shocked wind cooling}, we investigate the cooling physics of fast shocked winds, and show that the weak collisional coupling between protons and electrons suppresses the inverse Compton cooling rate significantly. We also show how mixing of the shocked wind with cooler gas modifies the effective cooling rate.  In \S \ref{outflow solutions}, we use the cooling rates derived in \S \ref{shocked wind cooling} to calculate representative outflow solutions. 
We show that the AGN-driven outflows are expected to be energy-conserving for a wide range of parameters relevant for observed systems. 
We argue that this conclusion is robust to the effects of mixing, and show how only modest confinement of the hot gas is necessary to produce momentum boosts comparable to those measured. 
We conclude in \S \ref{discussion} with a discussion of the implications of our analysis, in particular with respect to recent observations of galaxy-scale AGN winds. 

A series of appendices support the main text. 
Appendix \ref{self similar analysis} presents a general derivation of the conditions for energy and momentum conservation for self-similar outflow solutions. 
Appendix \ref{shocked wind emission} summarizes predictions for emission from shocked wind bubbles. 
Finally, Appendix \ref{numerical simulations} outlines how to incorporate the physics of AGN winds in numerical simulations.

\begin{figure}
\begin{center} 
\includegraphics[width=0.48\textwidth]{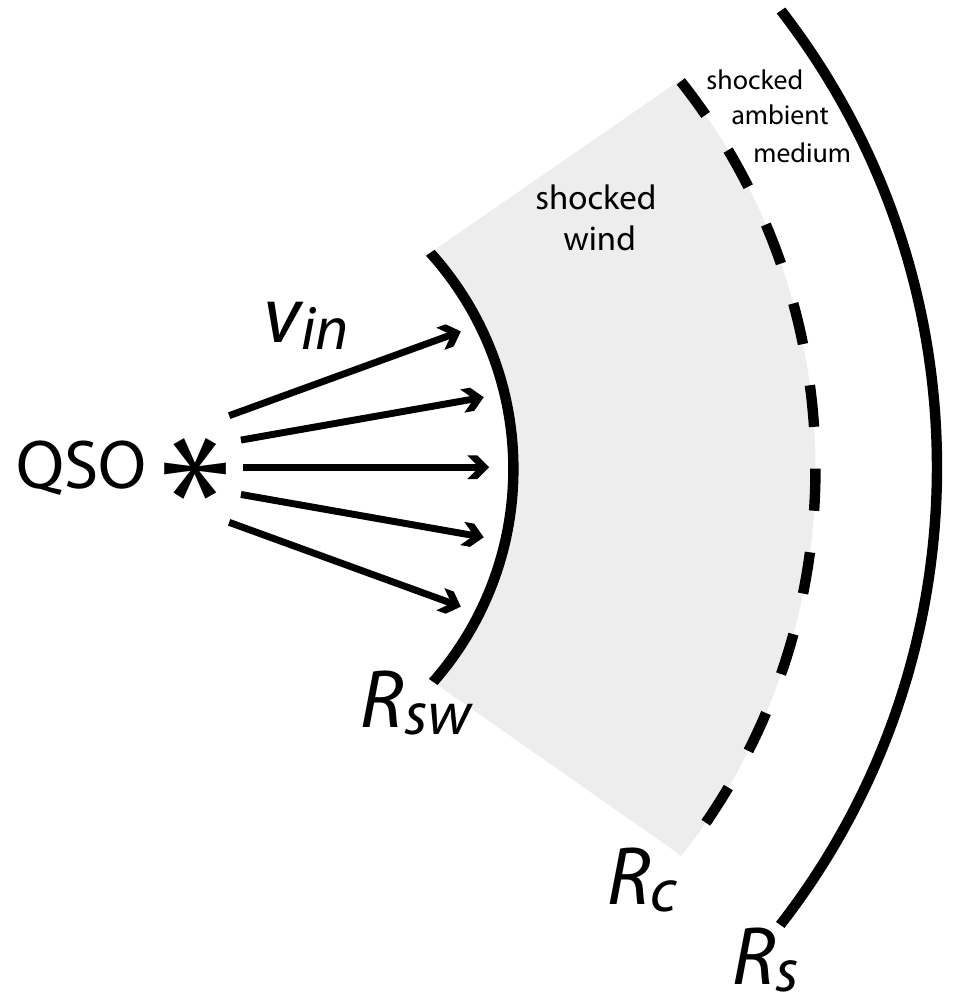}
\end{center}
\caption[]{Schematic illustration of the outflow structure. The AGN wind is launched in the galactic nucleus with velocity $v_{\rm in}$. 
The wind is shocked at $R_{\rm sw}$ (velocity $v_{\rm sw})$. A second shock, at radius $R_{\rm s}$ (velocity $v_{\rm s}$), is driven into the ambient ISM. The two shocks are separated by a contact discontinuity at $R_{\rm c}$. 
The swept-up ambient material piles up in a shell at $\approx R_{\rm s}$. However, it is the cooling of the shocked wind (not the shocked ambient medium) that determines whether the outflow conserves energy or momentum.
}
\label{wind structure figure} 
\end{figure}

\section{Cooling of the shocked wind}
\label{shocked wind cooling}
Our first goal is to derive the cooling rate of a fast, shocked AGN wind. 

\subsection{Problem setup and shock structure}
We focus first on the case of fast nuclear winds with launching speeds $v_{\rm in} \gtrsim 10,000$ km s$^{-1}$. 
For concreteness, we analyze the following idealized problem. At $t=0$, a central black hole with radiative luminosity $L_{\rm AGN}$ turns on and drives a constant, spherically-symmetric wind with initial velocity $v_{\rm in}$ and mass outflow rate $\dot{M}_{\rm in}$. 
For reference, the Eddington luminosity ($L_{\rm Edd}$) is related to the black hole mass ($M_{\rm BH}$) via
\begin{equation}
L_{\rm Edd} = \frac{4 \pi G M_{\rm BH} m_{\rm p} c}{\sigma_{\rm T}} \approx 1.3\times10^{46}~{\rm erg~s^{-1}} \left( \frac{M_{\rm BH}}{\rm 10^{8}~M_{\odot}} \right),
\end{equation}
where $G$ is Newton's constant, $m_{\rm p}$ is the mass of the proton, $c$ is the speed of light, and $\sigma_{\rm T}$ is the Thomson cross section. 
We also define the Eddington accretion rate $\dot{M}_{\rm Edd} \equiv L_{\rm Edd}/(\eta c^{2})$, where $\eta = 0.1$ is the radiative efficiency. 
The momentum flux of the small-scale wind is parameterized by
\begin{equation}
\dot{M}_{\rm in} v_{\rm in} \equiv \tau_{\rm in} \frac{L_{\rm AGN}}{c}.
\end{equation}
If $L_{\rm AGN} = L_{\rm Edd}$, then
\begin{equation}
\tau_{\rm in}  = \eta^{-1} \left( \frac{v_{\rm in}}{c} \right) \frac{\dot{M}_{\rm in}}{\dot{M}_{\rm Edd}}
\end{equation}
is related to the mass loading of the wind. 

The ambient, pre-shock medium is also assumed to be spherically symmetric, with a radial gas density profile $\rho_{\rm g}(R)$. 
The details of the small-scale wind acceleration are unspecified, but could owe to radiation pressure on spectral lines, free electrons, or dust, as well as hydro-magnetic forces (we return to the acceleration mechanisms in \S \ref{driving mechanisms}).  
We assume that this acceleration occurs on scales much smaller than all others in the problem. For instance, the accretion disk wind model of \cite{1995ApJ...451..498M} predicts that the wind is launched at a radius $\sim0.01$ pc.  
Since the interesting physics in our calculations occurs at radii where the escape velocity is $\ll v_{\rm in}$, we can also neglect gravity in our first-order calculations; the numerical calculations of \S \ref{outflow solutions} will include a more accurate treatment. 

Before proceeding, it is important to understand the structure of the outflow that develops. Studies of the analogous problem of fast stellar winds revealed that the outflow is characterized by both a forward and a reverse shock, separated by a contact discontinuity \citep[see Fig. \ref{wind structure figure}; e.g.,][]{1977ApJ...218..377W}. 
The outermost shock is the forward shock, which propagates in the ambient medium. The innermost, reverse shock is the shock first encountered by the gas accelerated by the BH; it is also termed the wind shock. 
We denote the radius of the wind shock by $R_{\rm sw}$, the radius of the contact discontinuity by $R_{\rm c}$, and the radius of the ambient medium shock by $R_{\rm s}$. 
$R_{\rm sw}$ and $R_{\rm s}$ have velocities $v_{\rm sw}$ and $v_{\rm s}$, respectively, with respect to the BH. 
As the outflow propagates outward, it sweeps up ambient ISM. 
This results in a shell of swept-up material at $\approx R_{\rm s}$ and causes the bubble to slow down its expansion. 
Except at very early times, we thus have $v_{\rm s} \ll v_{\rm in}$ and $v_{\rm sw} \ll v_{\rm in}$.\footnote{The last inequality may be transiently violated if the shocked wind quickly loses its pressure support and $R_{\rm sw} \to R_{\rm s}$ on a time scale as short as $\sim (R_{\rm s}-R_{\rm sw})/v_{\rm in}$, but this does not affect our arguments.} 
Then, the velocity of the wind shock relative to the wind itself is $v_{\rm ws} \equiv v_{\rm in} - v_{\rm sw} \approx v_{\rm in}$. 

Neglecting for the moment two-temperature effects, which we demonstrate are important in the next section, the temperature of the post-shock gas depends sensitively on the shock velocity. 
For a strong shock of velocity $v_{\rm sh}$ in a monatomic gas, the Rankine-Hugoniot jump conditions imply
\begin{equation}
\label{shock temperature}
T_{\rm sh}(v_{\rm sh})=\frac{3 \mu}{16k}m_{\rm p} v_{\rm sh}^{2}\approx 1.2\times10^{10}~{\rm K} \left( \frac{v_{\rm sh}}{\rm 30,000~km~s^{-1}} \right)^{2},
\end{equation}
where $\mu$ is the mean molecular weight. 
In general, $v_{\rm s} \ll v_{\rm in}$ and the shocked wind is much hotter than the shocked ambient medium. 
For instance, while the swept-up material may decelerate to a few 100 km s$^{-1}$ (or less), the velocity of the wind shock remains nearly constant at $v_{\rm ws}\approx v_{\rm in} \gtrsim 10,000$ km s$^{-1}$. 
This is key to the question of whether AGN outflows conserve energy or momentum, because it is much harder to cool the $T\gtrsim10^{9}$ K shocked wind than the cooler shocked ambient medium. 
In particular, \emph{it is the cooling of the shocked wind, rather than the shocked ambient medium, that determines whether the outflow is energy or momentum-conserving} \citep[see also][]{2011MNRAS.415L...6K}. 
Whether or not the shocked ambient medium cools has no significant impact on the global outflow dynamics, and for our purposes $R_{\rm c} \approx R_{\rm s}$ \citep[e.g.,][]{1992ApJ...388...93K}. We thus make the identification $R_{\rm c} \to R_{\rm s}$ in the following. 

To determine the cooling rate of the shocked wind, we need its characteristic density, temperature, and radius from the BH. 
Although the exact structure of the wind bubble is non-trivial, the analysis is considerably simplified, yet retains the essential physics, by adopting the following approximation. 
Since the sound crossing time of the shocked wind ($t_{\rm s} \sim [R_{\rm s} - R_{\rm sw}]/v_{\rm in}$) is short, the shocked wind pressure is nearly uniform between $R_{\rm sw}$ and $R_{\rm s}$. 
Furthermore, \cite{1992ApJ...388..103K} show that the gas density $\rho$ is approximately constant in the intermediate region $R_{\rm sw} \ll R \ll R_{\rm s}$, a result supported by the numerical calculations of \cite{1975A&A....43..323F}. 
Briefly, this follows because mass conservation and adiabaticity imply that $R^{2} v$ is constant in that region. Since the flow is steady, $\rho R^{2} v$ is also constant, implying that $\rho$ must be constant. 
As the pressure $P \propto \rho T$ is uniform, $T$ is approximately uniform as well. 
The properties of the shocked wind are thus adequately described by constants $\bar{\rho}_{\rm sw}(t)$, $\bar{T}_{\rm sw}(t)$, and $\bar{P}_{\rm sw}(t)$, where the density and temperature are given by the shock jump conditions at $R_{\rm sw}$. 
In particular,
\begin{equation}
\label{shocked wind density}
\bar{\rho}_{\rm sw} = 4 \rho_{\rm w}(R_{\rm sw}) = \frac{\dot{M}_{\rm in}}{\pi R_{\rm sw}^{2} v_{\rm in}}. 
\end{equation}
Before cooling of the shocked wind becomes important, the thermal energy of the wind bubble is conserved, $E_{\rm b} = (1/2) L_{\rm in} t$, where we assume that 1/2 of the energy goes into bulk motion of the swept up shell, 1/2 goes into thermal pressure of the shocked wind bubble,\footnote{This is a simple approximation to more exact calculations; \cite{1977ApJ...218..377W} show that for adiabatic bubbles in a uniform medium $E_{\rm b} \approx (5/11) L_{\rm in} t$.} and $L_{\rm in} \equiv (1/2) \dot{M}_{\rm in} v_{\rm in}^{2}$. 
The thermal pressure of the shocked wind is related to $E_{\rm b}$ via $\bar{P}_{\rm sw} = 2 E_{\rm b}/ 3 V_{\rm b}$, where $V_{\rm b} \equiv 4 \pi (R_{\rm s}^{3} - R_{\rm sw}^{3})/3$. 
Balancing the ram pressure from the unshocked wind with the thermal pressure of the shocked wind bubble at $R_{\rm sw}$, we obtain
\begin{equation}
\label{Rsw}
R_{\rm sw} = \left[ \frac{L_{\rm in}}{E_{\rm b} v_{\rm in}} (R_{\rm s}^{3} - R_{\rm sw}^{3}) \right]^{1/2} 
\end{equation}
When $R_{\rm sw} \ll R_{\rm s}$, which applies when the wind shock is not radiative, $R_{\rm sw} \sim R_{\rm s} (v_{\rm s}/v_{\rm in})^{1/2}$ (using $t \sim R_{\rm s}/v_{\rm s}$). 

Finally, the cooling of a Lagrangian element of shocked wind occurs principally at $\sim R_{\rm s}$, so this is the relevant radius to evaluate the inverse Compton cooling time below. 

\subsection{Cooling of the two-temperature plasma}
\label{cooling processes}
The key issue in determining the outflow type is whether the shocked wind cools. 
We distinguish between three regimes: energy-conserving, momentum-conserving, and the intermediate partially radiative bubble stage. 
The limit realized depends on the ordering of three time scales: the flow time $t_{\rm flow} \equiv R_{\rm s}/v_{\rm s}$, the crossing time $t_{\rm cr} \equiv R_{\rm sw}/v_{\rm in}$, and the cooling time of the shocked wind $t_{\rm cool}$ \citep[][]{1992ApJ...388..103K}. 
If $t_{\rm cool} \gg t_{\rm flow} > t_{\rm cr}$, then the outflow is energy conserving. 
If $t_{\rm cool} \ll t_{\rm cr} < t_{\rm flow}$, the outflow conserves momentum. 
In the intermediate partially radiative bubble stage $t_{\rm cr} \lesssim t_{\rm cool} \lesssim t_{\rm flow}$, most of the shocked wind cools and is compressed in a thin shell at $R_{\rm s}$. However the wind shock itself is non-radiative and most of the volume is filled by the portion of the wind that has not had time to cool yet. 
In this stage, the outflow conserves neither momentum nor energy. 

For fast winds, free free emission and inverse Compton scattering in the AGN radiation field are the most important cooling processes (e.g., King 2003; Ciotti \& Ostriker 1997, 2001 also emphasized the importance of Compton scattering for AGN inflows and outflows)\nocite{2003ApJ...596L..27K, 1997ApJ...487L.105C, 2001ApJ...551..131C}. 
In practice, free free cooling is subdominant at the high temperatures $T\gtrsim10^{9}$ K of shocked winds with $v_{\rm in}\gtrsim 10,000$ km s$^{-1}$, but we include it in our calculations for completeness. 

A complication neglected so far in the context of AGN winds concerns the coupling between protons and electrons in the shocked wind. 
This is important because most of the cooling occurs via the electrons, but the protons carry most of the thermal pressure. 
Observations of supernova remnants in fact indicate that the electrons and protons can be significantly decoupled after fast shocks \citep[e.g.,][]{2007ApJ...654L..69G}, as is also the case in the solar wind \citep[e.g.,][]{1988JGR....9312923S}. 
If the protons in shocked AGN winds are not well coupled to the electrons, then their thermal energy can be trapped in the wind, in which case the outflow effectively conserves energy. 
Such a situation is analogous to low radiative efficiency accretion flows, in which two-temperature plasmas are invoked as a means to prevent the energy from being radiated away \citep[e.g.,][]{1982Natur.295...17R, 1995ApJ...452..710N, 2001ASPC..224...71Q}.

We first discuss in more detail the empirical and theoretical evidence regarding two-temperature effects. Some measurements of Balmer-dominated SNR shocks indicate that $T_{\rm e}$ can be as low as the minimal heating case $\sim (m_{\rm e}/m_{\rm p}) T_{\rm p}$ for fast shocks with $v_{\rm sh} \gg 2,000$ km s$^{-1}$ \citep[e.g.,][]{2007ApJ...654L..69G}. However, the evidence is ambiguous for SNR shocks in general. For example, observations of the reverse shock in SN 1987A suggest an electron-proton temperature ratio as high as $T_{\rm e}/T_{\rm p} \approx 0.14-0.35$ \citep[][]{2011ApJ...743..186F}. 
In the solar wind, the faster shocks are measured to heat electrons to less than $10\%$ of the proton temperature \citep[][]{1988JGR....9312923S}. 
\cite{2006ESASP.604..723M} presented evidence for electron heating faster than predicted by Coulomb collisions in the bow shock of the ``Bullet'' galaxy cluster. 
However, the Mach number $\mathcal{M} \approx 3$ of the bow shock is much lower than expected for AGN winds. 
While electron heating is not well understood theoretically, particle-in-cell simulation demonstrate that electrons can be quickly heated to $T_{\rm e} \gtrsim 0.1 T_{\rm p}$ by plasma instabilities in collisionless shocks \citep[][]{2011ApJ...733...63R}. 
Importantly, though, the same simulations show that magnetic turbulence decays rapidly in the post-shock region \cite[][]{2008ApJ...674..378C}. 
As we show below, our results do not depend significantly on the level of ``prompt'' electron heating at the shock, but rather on the assumption that Coulomb collisions dominate afterward. 
We also note that even if some residual collisionless heating persisted well past the shock itself, the ratios $T_{\rm p}/T_{\rm e} \sim 10$ that we find below assuming Coulomb collisions only is comparable to (or less than) the ratios observationally determined in most fast shocks in SNRs and in the solar wind. 
Our calculations are thus relatively robust to uncertainties in collisionless processes and are well motivated by the available theoretical and empirical data.

\begin{figure}
\begin{center} 
\includegraphics[width=0.47\textwidth]{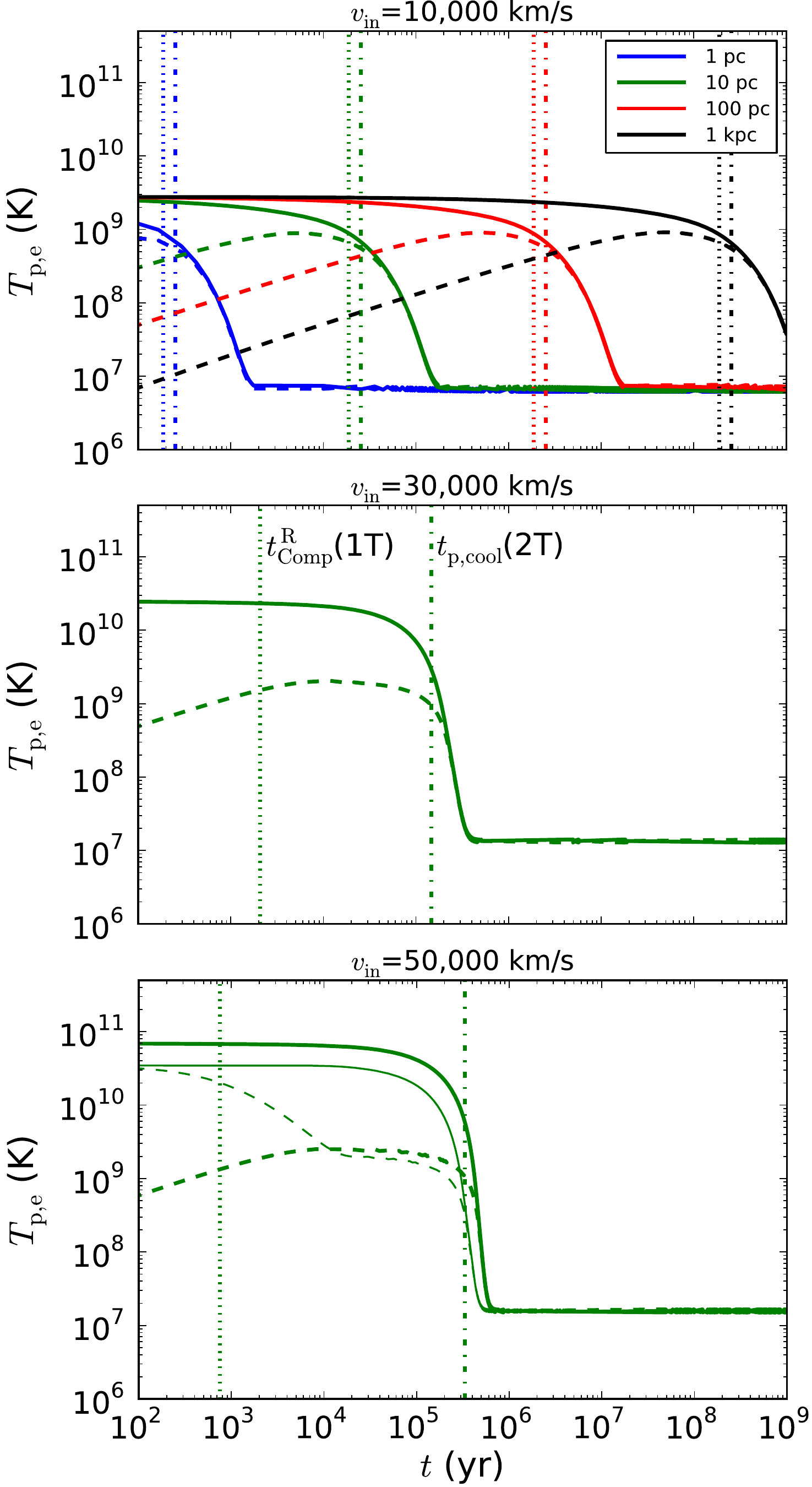}
\end{center}
\caption[]{Proton (solid) and electron (dashed) temperatures behind the wind shock, as a function of time since passing through the shock. Fiducial AGN parameters ($L_{\rm AGN}=10^{46}$ erg s$^{-1}$, $\tau_{\rm in}=1$), a swept-up shell velocity $v_{\rm s}=1,000$ km s$^{-1}$, and a Compton temperature $T_{\rm X}=2\times10^{7}$ K are assumed in all cases. The different colors in the top panel correspond to different radii of the swept-up shell ($R_{\rm s}$, assumed fixed), and the three panels show the results for different wind shock velocities $v_{\rm ws}\approx v_{\rm in}$. 
The shocked wind number density is given by equation (\ref{shocked wind density}). 
For reference, the vertical dotted lines show the relativistic inverse Compton cooling time evaluated assuming equipartition ($T_{\rm e} = T_{\rm sh}(v_{\rm in})$) in each case. The vertical dash-dotted lines show the effective proton cooling time taking into account two-temperature plasma effects (eq. (\ref{tp cool})).
This effective proton cooling time is significantly longer than the standard inverse Compton cooling time, especially at high $v_{\rm in}$ when the electrons would be relativistic in the absence of two-temperature effects. 
Most curves assume minimal electron heating at the shock ($T_{\rm e}=(m_{\rm e}/m_{\rm p})T_{\rm p}$). 
The thinner curves in the bottom panel show an example assuming full electron heating at the shock ($T_{\rm p} = T_{\rm e}$), but Coulomb coupling only afterward. 
The effective proton cooling time is not sensitive to the initial electron temperature, as long as Coulomb collisions dominate past the shock.
}
\label{TpTe figure} 
\end{figure}

Since most of the pre-shock kinetic energy is carried by the protons, minimal electron heating implies that those particles initially receive the majority of the shock heat, $T_{\rm p} \approx 2T_{\rm sh}(v_{\rm in})$, with electrons initially heated to $T_{\rm e} \sim T_{\rm p} (m_{\rm e}/m_{\rm p})$. As mentioned above, it is possible that plasma instabilities near the shock heat the electrons to higher temperatures. Our arguments are however not sensitive to the electron temperature immediately past the shock, as long as coupling in the post-shock region is dominated by Coulomb collisions. This is shown explicitly below, in Figure \ref{TpTe figure}, where we explore the case of complete heating at the shock, $T_{\rm p}=T_{\rm e}$. 
Assuming that Coulomb collisions in fact dominate in the post-shock region, 
electrons and protons achieve equipartition ($T_{\rm e} \approx T_{\rm p}$) on a time scale 
\begin{align}
\label{t_ei}
t_{\rm ei} & = \frac{3 m_{\rm e} m_{\rm p}}{8(2 \pi)^{1/2} n_{\rm p} e^{4} \ln{\Lambda}} \left( \frac{kT_{\rm e}}{m_{\rm e}} + \frac{kT_{\rm p}}{m_{\rm p}} \right)^{3/2},
\end{align}
where
\begin{equation}
\ln{\Lambda} \approx 39 + \ln{\left( \frac{T_{\rm e}}{\rm 10^{10}~K} \right)} - \frac{1}{2} \ln{\left( \frac{n_{\rm e}}{\rm 1~cm^{-3}} \right)} 
\end{equation}
\citep[][]{1962pfig.book.....S}.

Equation (\ref{t_ei}) assumes a fully ionized plasma consisting of protons and electrons only, an approximation which we adopt throughout. 
We fiducially assume $\ln{\Lambda}=40$ in this work. 
In our treatment of the shock, we further assume that there is a magnetic field sufficiently strong to suppress thermal conduction, but small enough to have a negligible dynamical impact. This allows us to treat electron-ion equilibrium locally. 
In this section, we use expressions appropriate in the limit of non-relativistic particles, which allow us to derive simple analytic approximations. 
It is not \emph{a priori} clear that this is a valid assumption since, for example, the electrons could be heated to relativistic speeds in reality. 
We have implemented an alternative set of equations using the expressions from \cite{1995ApJ...452..710N}, which are valid in both the relativistic and non-relativistic regimes, and confirmed that the results obtained here are accurate. 

The thermal free free emissivity is
\begin{equation}
\epsilon_{\rm ff}(T_{\rm e}) = 1.4 \times 10^{-27} T_{\rm e}^{1/2} C n_{\rm e}^{2} \bar{g}_{B},
\end{equation}
where all terms are in cgs units, $\bar{g}_{B}$ is the average Gaunt factor, and $C$ is a clumping factor (which we both set to 1 fiducially). 
The rate of energy transfer between the electrons and photons owing to Compton scattering is
\begin{align}
\frac{dU_{\rm ph}}{dt} = 4 U_{\rm ph} 
\sigma_{\rm T} n_{\rm e} c & \left( 
\frac{kT_{\rm e}}{m_{\rm e} c^{2}} - \frac{kT_{\rm X}}{m_{\rm e} c^{2}}  
\right) \\ \notag
& \times \left( 1 + \frac{5}{2} \frac{kT_{\rm e}}{m_{\rm e} c^{2}} - 2 \pi^{2} \frac{kT_{\rm X}}{m_{\rm e} c^{2} } \right),
\end{align}
where $T_{\rm X}$ Compton temperature of the radiation field \citep[e.g.,][]{2001AstL...27..481S} and
\begin{equation}
\label{Uph}
U_{\rm ph} = \frac{L_{\rm AGN}}{4 \pi R^{2} c}
\end{equation} 
is the radiation energy density. 

Let $U_{\rm e}$ and $U_{\rm p}$ be the electron and proton energy densities, respectively. For simplicity, we assume that the electron and proton distributions are individually thermal at all times, so that $U_{\rm e} = 3 n_{\rm e} k T_{\rm e}/2$ and $U_{\rm p} = 3 n_{\rm p} k T_{\rm p}/2$ (neutrality implies that $n_{\rm p} = n_{\rm e}$).  Assuming that the plasma is of uniform and constant density, the temperature evolution of the plasma is described by the following system of equations:
\begin{equation}
\frac{dT_{\rm e}}{dt} = - \frac{2}{3 k n_{\rm e}} \left[ \frac{dU_{\rm ph}}{dt} + \epsilon_{\rm ff}(T_{\rm e}) \right] + \frac{T_{\rm p} - T_{\rm e}}{t_{\rm ei}}
\end{equation}
\begin{equation}
\frac{dT_{\rm p}}{dt} = \frac{T_{\rm e} - T_{\rm p}}{t_{\rm ei}} .
\end{equation}
To develop an understanding of how the shocked wind cooling proceeds, we consider an idealized model in which $T_{\rm p} = 2 T_{\rm sh}(v_{\rm in})$, $T_{\rm e} = T_{\rm p} (m_{\rm e}/m_{\rm p})$ at $t=0$. 

Figure \ref{TpTe figure} shows the evolution of the proton and electron temperatures for the representative cases of $L_{\rm AGN}=10^{46}$ erg s$^{-1}$ (a $M_{\rm BH}=10^{8}$ M$_{\odot}$ black hole radiating near the Eddington limit), $v_{\rm in}=10,000,~30,000$ and 50,000 km s$^{-1}$, and $\tau_{\rm in}=1$. The different sets of curves in the top panel correspond to radii $R_{\rm s}=1,~10,~100,$ and 1,000 pc, from left to right. A Compton temperature $T_{\rm X}=2\times10^{7}$ K is assumed based on the average AGN spectrum from \cite{2004MNRAS.347..144S}. 
For these calculations, the density of the shocked wind is approximated by 
\begin{equation}
\label{shocked wind density}
n_{\rm p} = \frac{\dot{M}_{\rm in}}{\pi R_{\rm sw}^{2} v_{\rm in} m_{\rm p}} \approx \frac{\dot{M}_{\rm in}}{\pi R_{\rm s}^{2} v_{\rm s} m_{\rm p}} = \frac{\tau_{\rm in}L_{\rm AGN}}{\pi R_{\rm s}^{2} v_{\rm s} v_{\rm in} c m_{\rm p}}.
\end{equation}
The second equality assumes that the outflow is energy-conserving ($R_{\rm sw}\approx(v_{\rm s}/v_{\rm in})^{1/2} R_{\rm s}$). 
The calculations also assume an outer shock velocity $v_{\rm s}=1,000$ km s$^{-1}$. This is not necessarily self-consistent, as a given ambient density profile implies a relation between $R_{\rm s}$ and $v_{\rm s}$; self-consistent models are constructed in \S \ref{outflow solutions}. 

For reference, the vertical dotted lines show the relativistic inverse Compton cooling time evaluated assuming equipartition ($T_{\rm e} = T_{\rm sh}(v_{\rm in})$) in each case, 
\begin{align}
\label{Compton time rel}
t_{\rm Comp}^{\rm R} &= \frac{3 m_{\rm e}^{2} c^{3}}{32 \sigma_{\rm T} U_{\rm ph} k T_{\rm e}} \\ \notag
& \approx 2 \times 10^{7}~{\rm yr}~\left( \frac{R_{\rm s}}{\rm 1~kpc} \right)^{2} \left( \frac{L_{\rm AGN}}{\rm 10^{46}~erg~s^{-1}} \right)^{-1} \\ \notag
&~~~~~~~~~~~~~~~~~~~~~~~~~~~~~~ \times \left( \frac{v_{\rm in}}{\rm 30,000~km~s^{-1}} \right)^{-2}.
\end{align}
This is the relevant cooling limit neglecting two-temperature effects for $v_{\rm in} \gtrsim 10,000$ km s$^{-1}$ \citep[e.g., as in][]{2003ApJ...596L..27K}.

Particularly for $v_{\rm in}=30,000$ km s$^{-1}$ and $v_{\rm in}=50,000$ km s$^{-1}$, the protons take significantly longer than $t_{\rm Comp}^{\rm R}$ to lose half of their thermal energy. 
Moreover, the effective cooling time of the protons \emph{increases} with increasing $v_{\rm in}$, a trend \emph{opposite} to standard inverse Compton cooling for relativistic electrons (eq. (\ref{Compton time rel})). 
The electrons in fact remain slow enough that their inverse Compton scattering time scale follows the non-relativistic expression to a good approximation. 
In contrast, assuming $T_{\rm e} \sim T_{\rm p}$ would imply a root mean square (rms) electron velocity $\langle v_{\rm e}^{2} \rangle^{1/2} \approx 43 \langle v_{\rm p}^{2} \rangle^{1/2}$ (where $\langle v_{\rm p} \rangle^{1/2}$ is the proton rms velocity), putting the electrons into the relativistic limit for these shock velocities.  
Thus, the effective cooling time of the shocked protons is not the standard inverse Compton cooling time, but is longer owing to two-temperature plasma effects.

We now derive an analytic approximation to the effective cooling time of the shocked protons. 
First, it is apparent from Figure \ref{TpTe figure} that the electron temperature reaches a plateau before the protons cool significantly. 
This temporary ``equilibrium'' electron temperature, $T_{\rm e}^{\rm eq}$, is set by a balance between Coulomb heating by the protons and inverse Compton cooling in the AGN radiation field. 
Setting $dT_{\rm e}/dt=0$, neglecting free free cooling, approximating inverse Compton cooling by $dU_{\rm ph}/dt = (4 k T_{\rm e} / m_{\rm e} c) \sigma_{\rm T} n_{\rm e} U_{\rm ph}$ (exact for a thermal, non-relativistic electron distribution and $T_{\rm X} \ll T_{\rm e}$), and in the limit $T_{\rm e}^{\rm eq}/m_{\rm e} \gg T_{\rm p}/m_{\rm p}$, 
\begin{align}
\label{Te eq}
T_{\rm e}^{\rm eq} & \approx (2 \pi)^{1/5} 
\left[
\frac{m_{\rm e}^{3} e^{8} c^{2} (\ln{\Lambda})^{2} n_{\rm p}^{2} T_{\rm p}^{2}}{\sigma_{\rm T}^{2} k^{3} m_{\rm p}^{2} U_{\rm ph}^{2}}
\right]^{1/5}
\\ \notag
& \approx 3^{2/5} \left( \frac{\pi}{2} \right)^{1/5} 
\left[
\frac{m_{\rm e}^{3} \mu^{2} e^{8} c^{2} (\ln{\Lambda})^{2} v_{\rm in}^{2} \tau_{\rm in}^{2}}{\sigma_{\rm T}^{2} k^{5} m_{\rm p}^{2} v_{\rm s}^{2}}
\right]^{1/5}
 \\ \notag
& \approx 2.9\times10^{9}~{\rm K} \left( \frac{v_{\rm in}/v_{\rm s}}{30} \right)^{2/5}
\left( \frac{\ln{\Lambda}}{40} \right)^{2/5} \tau_{\rm in}^{2/5}.
\end{align}
The intermediate equality holds assuming $U_{\rm ph}$ and $n_{\rm p}$ are given by equations (\ref{Uph}) (with $R=R_{\rm s}$) and (\ref{shocked wind density}), respectively, and $T_{\rm p}= 2 T_{\rm sh}(v_{\rm in})$.

This analytic approximation is in excellent agreement with the numerical results in Figure \ref{TpTe figure} (for the $v_{\rm in}=10,000$ km s$^{-1}$ case, proton cooling is significant before the plateau is reached). 
Coulomb collisions with the electrons is the only process through which protons lose energy. The proton cooling time is therefore given by $t_{\rm p,cool} \approx t_{\rm ei}(T_{\rm e}^{\rm eq})$, i.e.,
\begin{align}
\label{tp cool}
t_{\rm p,cool} & \approx \frac{3^{8/5}}{8} \left( \frac{\pi}{2} \right)^{4/5} 
\left[ \frac{m_{\rm e}^{2} m_{\rm p}^{7} \mu^{3} v_{\rm s}^{2} v_{\rm in}^{8} c^{8}}{\sigma_{\rm T}^{3} e^{8} (\ln{\Lambda})^{2} \tau_{\rm in}^{2}} \right]^{1/5} \frac{R_{\rm s}^{2}}{L_{\rm AGN}} \\ \notag
& \approx 1.4\times10^{9}{\rm~yr} \left( \frac{R_{\rm s}}{\rm 1~kpc} \right)^{2} \left( \frac{L_{\rm AGN}}{\rm 10^{46}~erg~s^{-1}} \right)^{-1} \\ \notag
& \times \left( \frac{v_{\rm s}}{\rm~1,000~km~s^{-1}} \right)^{2/5} \left( \frac{v_{\rm in}}{\rm~30,000~km~s^{-1}} \right)^{8/5} \\ \notag
& \times \left( \frac{\ln{\Lambda}}{40} \right)^{-2/5} \tau_{\rm in}^{-2/5}.
\end{align}
Although this time scale derives from two-temperature effects, it is nearly independent of the electron temperature immediately past the shock. 
This is because $T_{\rm e}^{\rm eq}$ is set by a balance between inverse Compton cooling and Coulomb heating of the electrons in the post-shock region. In particular, independent of the initial temperature of the electrons at the shock, $T_{\rm e}^{\rm eq}$ approaches the value given by equation (\ref{Te eq}) in the downstream region (see Fig. \ref{TpTe figure}, bottom panel) and thus the estimate of the effective proton cooling time in equation (\ref{tp cool}) applies.

Finally, consider synchrotron cooling. For a given electron population, the ratio of the synchrotron to inverse Compton power is
\begin{equation}
\label{synch to Comp ratio}
\frac{P_{\rm synch}}{P_{\rm Comp}} = \frac{U_{\rm B}}{U_{\rm ph}},
\end{equation}
where $U_{\rm B} \equiv B^{2} / 8 \pi$ is the magnetic energy density. 
For an AGN radiation energy density following equation (\ref{Uph}), 
\begin{align}
\label{P synch}
\frac{P_{\rm synch}}{P_{\rm Comp}} &= \frac{B^{2} R^{2} c}{2 L_{\rm AGN}} \\ \notag
& \approx 1.4\times10^{-5} \left( \frac{B}{\rm 1~\mu G} \right)^{2} \left( \frac{R}{\rm 1~kpc} \right)^{2} \\ \notag
&~~~~~~~~~~~~~~~~~~~~~~~~~~~~~~\times \left( \frac{L_{\rm AGN}}{\rm 10^{46}~erg~s^{-1}} \right)^{-1}.
\end{align}
In ordinary spiral galaxies, $B\sim1-10$ $\mu$G \citep[][]{2001SSRv...99..243B} and synchrotron cooling is clearly subdominant. 
In starbursts, the magnetic field can reach $B\sim1$ mG \citep[][]{2006ApJ...645..186T}. 
Compact starbursts, such as local ULIRGs, however have gas disk scale heights $\sim 50$ pc \citep[e.g.,][]{1998ApJ...507..615D}. 
Provided that the magnetic field scale height is less than a few times this value, synchrotron cooling should also be less efficient than inverse Compton cooling in those objects. 
We thus neglect synchrotron cooling of the shocked wind. 

\subsection{Mixing and other cooling limits}
\label{mixing}
Equation (\ref{tp cool}) is the cooling time appropriate in the absence of mixing with ambient gas. 
Here, we study the effects of mixing of cold gas with the hot shocked wind in the limit in which the mixed cold and hot gas is rapidly homogenized. 
Then, the amount of mixing can be parameterized by the fraction
\begin{equation}
f_{\rm mix} \equiv \frac{M_{\rm sw}+M_{\rm cold}}{M_{\rm sw}},
\end{equation}
where $M_{\rm cold}$ is the total mass of cold gas that is mixed with the shocked wind, and $M_{\rm sw} \equiv \dot{M}_{\rm in} t$. 
$M_{\rm cold}$ can consist of contributions from cold clumps ablated by the wind and from mixing of the swept-up ambient medium owing to hydrodynamical instabilities. 

Consider first a wind with $v_{\rm in} \gtrsim 10,000$ km s$^{-1}$ which initially develops a two-temperature structure past the wind shock. 
Initially, the pressure support $\bar{P}_{\rm sw} \propto n_{\rm p} T_{\rm p}$ is preserved when mixing occurs.  
As shown in \S \ref{cooling processes}, $T_{\rm e}^{\rm eq}$ also depends only on the product $n_{\rm p} T_{\rm p}$, and is thus also unaffected by mixing. 
In particular, in the absence of efficient collisionless coupling in the shocked wind, the electrons remain non-relativistic. 
We term this the \emph{Coulomb limit}. 

On the other hand, the effective proton cooling time in the two-temperature plasma approximation,
\begin{equation}
\label{tp cool mixing}
t_{\rm p,cool} \approx t_{\rm ei}(T_{\rm e}^{\rm eq}) \propto \frac{1}{n_{\rm p}} \propto \frac{1}{f_{\rm mix}},
\end{equation}
is reduced linearly with the amount of mixing. 
When the time scale for energy exchange between the electrons and the protons becomes shorter than the inverse Compton scattering time, the latter process becomes the rate limiting step. 
We term this the \emph{inverse Compton cooling limit}. It is realized when the non-relativistic inverse Compton cooling time
\begin{align}
\label{Compton time nonrel}
t_{\rm Comp}^{\rm NR} & = 
\frac{3 \pi}{2} \frac{m_{\rm e} c^{2} R_{\rm s}^{2}}{\sigma_{\rm T} L_{\rm AGN}} \\ \notag
& \approx 2 \times 10^{8}~{\rm yr} \left( \frac{R_{\rm s}}{\rm 1~kpc} \right)^{2} \left( \frac{L_{\rm AGN}}{\rm 10^{46}~erg~s^{-1}} \right)^{-1}. 
\end{align} 
equals $t_{\rm p,cool}$, corresponding to
\begin{align}
f_{\rm mix}^{\rm Comp} &= \frac{3^{3/5}}{4 \pi^{1/5} 2^{4/5}} 
\left[
\frac{m_{\rm p}^{7} \sigma_{\rm T}^{2} \mu^{3} v_{\rm s}^{2} v_{\rm in}^{8}}{m_{\rm e}^{3} c^{2} e^{8} (ln{\Lambda})^{2} \tau_{\rm in}^{2}} \right]^{1/5} \\ \notag
& \approx 8.1 \left( \frac{v_{\rm s}}{\rm 1,000~km~s^{-1}} \right)^{2/5} \left( \frac{v_{\rm in}}{\rm 30,000~km~s^{-1}} \right)^{8/5} \\ \notag
&~~~~~~~~~~ \times \left( \frac{\ln{\Lambda}}{40} \right)^{-2/5} \tau_{\rm in}^{-2/5}.
\end{align}
Since the $t_{\rm Comp}^{\rm NR}$ is independent of density and temperature, it is constant as mixing proceeds. 

As the amount of mixing becomes larger, $f_{\rm mix} \to \infty$, the increase in the shocked wind bubble density and the decrease in its interior temperature both shorten the free free cooling time,
\begin{equation}
\label{t ff}
t_{\rm ff} \equiv \frac{3 n_{\rm e} k T_{\rm e}/2}{\epsilon_{\rm ff}}
\approx \frac{\rm 4.7\times10^{6}~yr}{\bar{g}_{\rm B} C}
 \left( \frac{T_{\rm e}}{\rm~10^{6}~K} \right)^{1/2} \left( \frac{n_{\rm e}}{\rm 1~cm^{-3}} \right)^{-1}.
\end{equation}
Multiplying equation (\ref{shocked wind density}) by $f_{\rm mix}$ to estimate the shocked wind density after mixing (assuming energy conservation) and using $T_{\rm e} = T_{\rm sh}(v_{\rm in}) / f_{\rm mix}$, we solve for $f_{\rm mix}$ such that $t_{\rm ff} = t_{\rm Comp}^{\rm NR}$:
\begin{align}
f_{\rm mix}^{\rm ff} & \approx \frac{26.4}{(C \bar{g}_{\rm B} \tau_{\rm in})^{2/3}} \left( \frac{v_{\rm s}}{\rm 1,000~km~s^{-1}} \right)^{2/3} \\ \notag
&~~~~~~~~~~~~~~~~~~~~~~~~~~ \times \left( \frac{v_{\rm in}}{\rm 30,000~km~s^{-1}} \right)^{4/3}.
\end{align}
In this \emph{free free limit}, $t_{\rm ff} \propto f_{\rm mix}^{-3/2}$.

It is important to note that the $f_{\rm mix}$ thresholds derived above determine the rate limiting step for cooling but do not imply that catastrophic cooling occurs. 
We show in \S \ref{outflow solutions} that substantially more mixing is needed for the shocked wind to radiate away its energy in a flow time in realistic conditions. 

For slower winds, $f_{\rm mix}^{\rm Comp}$ and $f_{\rm mix}^{\rm ff}$ can be $<1$.  This simply implies that the cooling of the shocked bubble may never be in the Coulomb or inverse Compton limits. For example, if both $f_{\rm mix}^{\rm Comp}<1$ and $f_{\rm mix}^{\rm ff}<1$, then the shocked wind cooling is dominated by free free emission even in the absence of mixing. This is the case for $v_{\rm in} \sim 1,000$ km s$^{-1}$.  For $v_{\rm in} \lesssim 1,000$ km s$^{-1}$, bound-free emission dominates over free free; we do not consider this limit further here.

At each radius, there is a physical upper limit to $f_{\rm mix}$, realized when all the swept-up gas mass has been mixed with shocked wind. Let $M_{\rm s}$ be the swept-up gas mass at $R_{\rm s}$, defined as the gas mass originally enclosed within that radius. 
Then, the maximum mixing parameter at $R_{\rm s}$ is
\begin{equation}
\label{fmix max}
f_{\rm mix}^{\rm max}(R_{\rm s}) = \frac{M_{\rm sw} + M_{\rm s}}{M_{\rm sw}}.
\end{equation}
In our numerical integrations (\S \ref{outflow solutions}), the mixing parameter at each radius is thus taken to be the minimum of the prescribed value and $f_{\rm mix}^{\rm max}(R_{\rm s})$.

\section{Outflow Solutions}
\label{outflow solutions}
We now use the results of the previous section on the cooling rate of the shocked wind to calculate outflow solutions. 
We focus here on numerical examples with parameters representative of AGN systems, but a general analysis of self-similar solutions in Appendix \ref{self similar analysis} demonstrates that our conclusions regarding energy conservation apply more generically. In what follows, the gas density profile of the ambient medium is parameterized by
\begin{equation}
\label{power law density}
\rho_{\rm g}(R) = \rho_{0} \left( \frac{R}{R_{0}} \right)^{-\alpha}.
\end{equation}
We use $R_{0}=100$ pc for all the numerical calculations in this work.

\subsection{Thin shell approximation}
\label{thin shell approx}
Our approach is similar to that of \cite{1977ApJ...218..377W}, but we include additional cooling processes, gravity, and approximate the effects of mixing following \S \ref{mixing}. 
We assume spherical symmetry and that conduction is negligible. 

The basic equation of motion is
\begin{equation}
\label{weaver eom}
\frac{d}{dt}(M_{\rm s} v_{\rm s}) = 4\pi R_{\rm s}^{2}(P_{\rm b} - P_{0}) - \frac{G M_{\rm s}M_{\rm t}}{R_{\rm s}^{2}},
\end{equation}
where $P_{\rm b}$ is the thermal pressure in the shocked wind bubble, $P_{\rm 0}$ is the pressure of the ambient medium, $M_{\rm s} = \int_{0}^{R_{\rm s}} dV \rho_{\rm g}$ is the small of the swept-up gas shell, and $M_{\rm t}=M_{\rm BH} + M_{\rm gal}(<R_{\rm s})$ accounts for the total gravitational mass within $R_{\rm s}$, including the central massive black hole and the surrounding galaxy. 
The gravitational potential of the galaxy is assumed to follow an isothermal sphere with velocity dispersion $\sigma$, 
\begin{align}
M_{\rm gal}(<R) = \frac{2 \sigma^{2} R}{G}. 
\end{align}
In the calculations that follow, we fix this total potential even when varying the properties of the ambient gas density profile (eq. (\ref{power law density})). 
We do this because, as we argue in the next section, the effective gas density swept up by the outflow is not necessarily directly tied to the total gravitational mass. 
The velocity dispersion is set to $\sigma=200$ km s$^{-1}$, appropriate for $M_{\rm BH}=10^{8}$ $M_{\odot}$ black holes on the $M_{\rm BH}-\sigma$ relation \citep[e.g.,][]{2009ApJ...698..198G}. 
This allows us to explore the dependences on the properties of the ambient medium while keeping the gravitational potential fixed. 
We have verified that instead assuming a total potential dominated by the gas mass does not change our results significantly. 

The thermal energy in the shocked wind region, $E_{\rm b}$, is related to its pressure, $P_{\rm b}$, by
\begin{equation}
E_{\rm b} = 2 \pi P_{\rm b} (R_{\rm s}^{3} - R_{\rm sw}^{3}).
\end{equation}
The location of the wind shock, $R_{\rm sw}$, is given by pressure balance as in equation (\ref{Rsw}). 
The rate of change of the energy in the shocked wind is determined by the energy injection rate, work done on the swept-up ISM, and cooling losses:
\begin{equation}
\dot{E_{\rm b}} = \frac{1}{2} \dot{M}_{\rm in} v_{\rm in}^{2} - 4 \pi R_{\rm s}^{2} P_{\rm b} \dot{R}_{\rm s} - L_{\rm b}, 
\end{equation}
where $L_{\rm b}$ is the radiative cooling rate. 
The appropriate cooling limit depends on the amount of mixing, parameterized by $f_{\rm mix}$ (\S \ref{mixing}). We define it as $L_{\rm 2T}$, $L_{\rm Comp}$, or $L_{\rm ff}$, depending on whether $f_{\rm mix}<f_{\rm mix}^{\rm Comp}$, $f_{\rm mix}^{\rm Comp} \leq f_{\rm mix} < f_{\rm mix}^{\rm ff}$, or $f_{\rm mix} \geq f_{\rm mix}^{\rm ff}$, where:
\begin{equation}
\label{sw luminosities}
L_{\rm 2T} = \frac{\mu E_{\rm b}}{t_{\rm p,cool}};~~~~~L_{\rm Comp} = \frac{\mu E_{\rm b}}{t_{\rm Comp}^{\rm NR}};~~~~~L_{\rm ff} = \frac{\mu E_{\rm b}}{t_{\rm ff}}.
\end{equation}
At each radius, $f_{\rm mix}$ is limited by $f_{\rm mix}^{\rm max}(R_{\rm s})$ (eq. (\ref{fmix max})). 
We do not keep track of the electron and proton temperatures separately, but instead use the approximation in equation (\ref{tp cool}) for $t_{\rm p,cool}$. 
We will also explore cases in which we neglect two-temperature effects on the cooling rate of the shocked wind. For those, $t_{\rm p,cool}$ is replaced by $t_{\rm Comp}^{\rm R}$ (eq. (\ref{Compton time rel})). 
Note that the mean molecular weight pre-factors are included in equation (\ref{sw luminosities}) because the cooling time scales were defined relative to the thermal energy carried by the electrons (or protons) only.

\begin{figure}
\begin{center}
\mbox{
\includegraphics[width=0.47\textwidth]{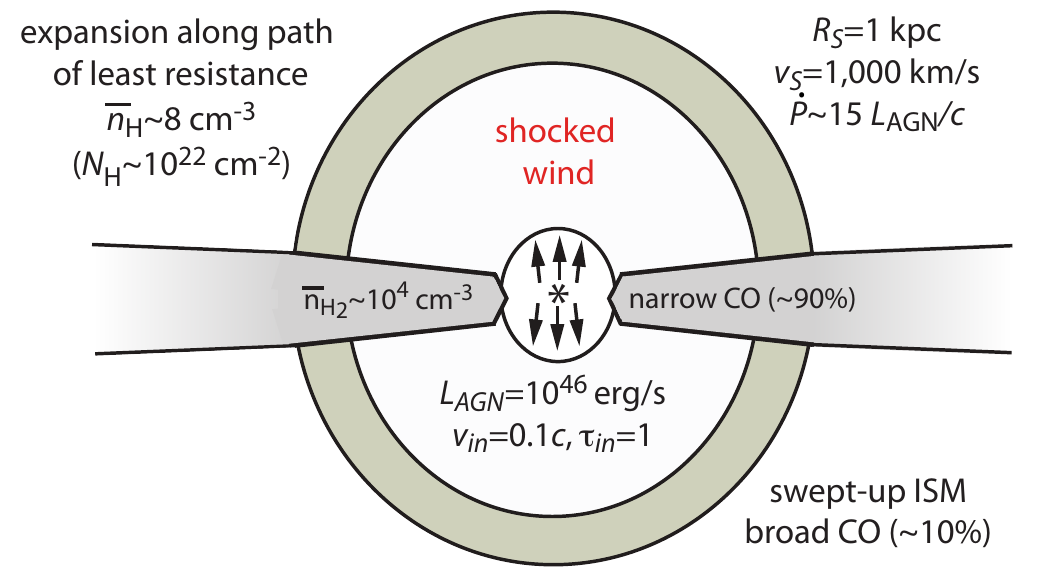}
}
\end{center}
\caption[]{Example of an outflow escaping preferentially along paths of least resistance above and below the dense molecular disk in a ULIRG system. Even though the gas density along the directions of escape is $<10^{-3}$ that of the molecular disk, the momentum boost can be boosted by factor $\sim 15$ during the expansion to a kiloparsec.}
\label{confinement example} 
\end{figure}

\begin{figure*}
\begin{center}
\includegraphics[width=0.9\textwidth]{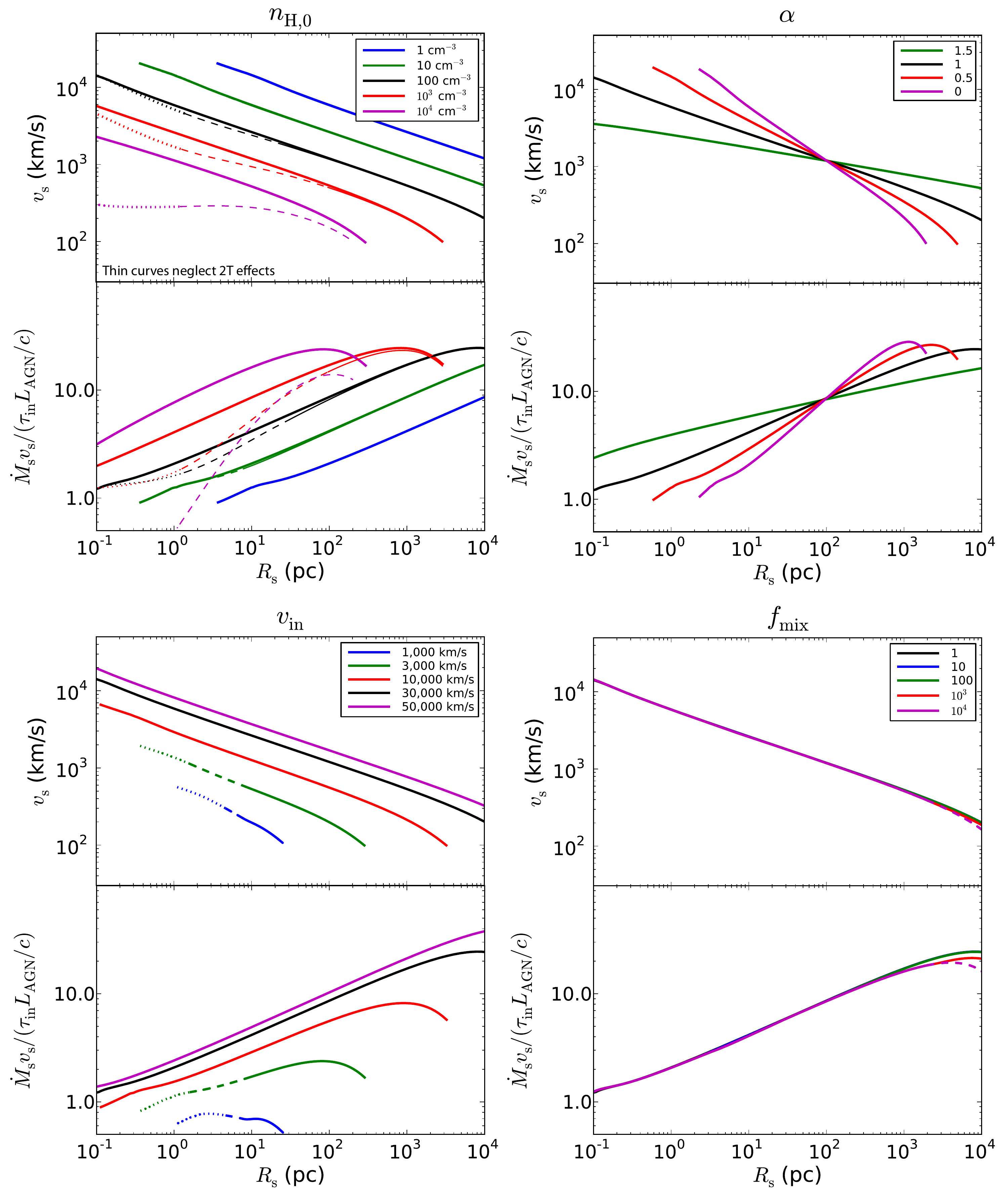}
\end{center}
\caption[]{Dependence of the outflow solution on $n_{\rm H,0}$ (density at $R_{0}=100$ pc), $\alpha$ (where $n_{\rm H} \propto R^{-\alpha}$), $v_{\rm in}$, and $f_{\rm mix}$. The fiducial parameters, motivated by local ULIRGs, are: $L_{\rm AGN}=10^{46}$ erg s$^{-1}$, $f_{\rm mix}=1$, $\alpha=1$, $n_{\rm H,0}=100$ cm$^{-3}$, $v_{\rm in}=30,000$ km s$^{-1}$, and $\tau_{\rm in}=1$. In all cases, the total gravitational potential is described by an isothermal sphere with $\sigma=200$ km s$^{-1}$, but the effective ambient gas density profile is varied.   
The solid line segments indicate periods during which the outflow is energy-conserving ($t_{\rm cool} > t_{\rm flow}$), the dashed segments indicate periods during which the outflow is in a partially radiative bubble stage ($t_{\rm cr} < t_{\rm cool} < t_{\rm flow}$), and the dotted segments indicate periods during which the outflow is momentum-conserving. In each case, the curve starts at the free expansion radius, $R_{\rm f}$ (see Appendix \ref{self similar analysis}). 
The thin curves in the top left panel illustrate the impact of neglecting two-temperature effects on inverse Compton cooling. 
} 
\label{numsols}
\end{figure*}

\subsection{Escape along paths of least resistance}
\label{leakage}
For a galaxy-scale outflow with momentum $P = \tau L_{\rm AGN} t_{\rm flow} / c$ (the normalization $\tau$ is to be distinguished from the parameter $\tau_{\rm in}$ characterizing the nuclear wind), momentum conservation implies that the average enclosed mass density is, assuming spherical symmetry and neglecting gravity,
\begin{align}
\label{momentum constraint}
\bar{n}_{\rm H}(<R_{\rm s}) & \approx \frac{3 \tau}{4 \pi R_{\rm s}^{2} c m_{\rm p}} \frac{L_{\rm AGN}}{v_{\rm s}^{2}} \\ \notag
& \approx 0.5~{\rm cm}^{-3}~\tau~\left( \frac{L_{\rm AGN}}{\rm 10^{46}~erg~s^{-1}} \right) 
\left( \frac{R_{\rm s}}{\rm 1~kpc} \right)^{-2} \\ \notag
& ~~~~~~~~~~~~~~~~~~~~~~~~~~~~~~~\times \left( \frac{v_{\rm s}}{\rm 1,000~km~s^{-1}} \right)^{-2}.
\end{align}
Here, we use $n_{\rm H}$ to denote the total hydrogen number density, even if it is in molecular form. 
In ULIRGs with observed outflows, this is much lower than the mean densities $\bar{n}_{\rm H}\sim10^{3}-10^{4}$ cm$^{-3}$ of the dense molecular disks on scales $\sim 500$ pc \citep[e.g.,][]{1998ApJ...507..615D}.\footnote{The dense molecular disks can extend to a radii $>1$ kpc. The host galaxies (including the stellar disks and tidal features) are typically significantly more extended.} 
Thus, the observed galaxy-scale outflows in these systems must have propagated along paths of least resistance. 
In the prototypical example Mrk 231, the neutral outflow extends to $R_{\rm s} \sim 3$ kpc and has $v_{\rm s} \sim 1,000$ km s$^{-1}$ \citep[][]{2011ApJ...729L..27R}. 
As this system is viewed nearly face-on, the outflow is most likely observed as it escapes approximately normal to the dense molecular disk. 

To illustrate how little confinement is necessary to produce a significant momentum boost, consider the example shown in Figure \ref{confinement example}, based on the fiducial AGN parameters used previously. 
In this case, the shocked wind escapes predominantly above and below a dense molecular disk. 
The molecular disks in ULIRGs have scale radii $R_{\rm d} \sim500$ pc and scale heights $h_{\rm d}\sim 50$ pc \citep[e.g.,][]{1998ApJ...507..615D}. 
The inward gravitational force on such a disk is
\begin{equation}
F_{\rm d,grav} \approx \frac{G M_{\rm d} M_{\rm t}(<R_{\rm d})}{R_{\rm d}^{2}}.
\end{equation}
Assuming that the AGN outflow is isotropic before encountering the disk, the outward force it exerts is
\begin{equation}
F_{\rm d,out} \approx \dot{P} \left( \frac{h_{\rm d}}{R_{\rm d}} \right),
\end{equation}
where $h_{\rm d}/R_{\rm d}$ is the fraction of the momentum flux intercepted by the disk. 
Thus,
\begin{align}
\label{outward to gravity force}
\frac{F_{\rm d,out}}{F_{\rm d,grav}} & \approx \tau \frac{L_{\rm AGN}}{c} \frac{h_{\rm d} R_{\rm d}}{G M_{\rm d} M_{\rm t}(<R_{\rm d})} \\ \notag
& \approx 0.07 \tau \left( \frac{L_{\rm AGN}}{\rm 10^{46}~erg~s^{-1}} \right) \left( \frac{h_{\rm d}}{\rm 50~pc} \right) \left( \frac{R_{\rm d}}{\rm 500~pc} \right) \\ \notag
& ~~~~~~~~~~ \times \left( \frac{M_{\rm d}}{\rm~10^{9}~M_{\odot}} \right)^{-1} \left( \frac{M_{\rm t}(<R_{\rm d})}{\rm~5\times10^{9}~M_{\odot}} \right)^{-1}.
\end{align}
Therefore, AGN outflows generally cannot eject the massive molecular disks in those systems in an instantaneous sense, unless $\tau \gg 1$. 
This is consistent with CO observations of ULIRGs, in which the outflowing gas with $v_{\rm s} \sim 1,000$ km s$^{-1}$ contributes only a small fraction $\sim10\%$ of the integrated CO luminosity \citep[][]{2010A&A...518L.155F}, most of the flux being concentrated in a narrow component identified with the disk. 
On the other hand, $\tau \gg 1$ can be achieved if there is sufficient confinement of the hot gas. 
Indeed, \cite{2011ApJ...732....9G} observe significant non-gravitational disturbances over the entire host galaxies of luminous obscured quasars at $z < 0.5$. 
Regardless, the wind is generally expected to preferentially escape normal to the disk. 

Even though the mean ambient density normal to the disk is $<10^{-3}$ of the molecular disk density in this example, the kiloparsec-scale outflow still reaches a momentum flux $\dot{P} \sim 15 L_{\rm AGN}/c$ (see \ref{momentum boost sec} below). Note that the implied column density $N_{\rm H} \sim R_{\rm s} \bar{n}_{\rm H} \sim 10^{22}$ cm$^{-3}$ is relatively modest, especially if we consider the column \emph{before} feedback provided by the wind exposes the AGN \citep[e.g.,][]{1999ApJ...522..157R, 2009ApJ...696..110T}. 
In reality, the outflow propagation depends on the details of the multiphase structure of the ambient ISM, and this aspect will be best addressed using three-dimensional simulations. 
Nevertheless, this demonstrates that escape along paths of least resistance does not preclude large momentum boosts, unless these paths are extremely under-dense. 

\subsection{Representative examples}
In Figure \ref{numsols}, we illustrate the dependence of thin shell outflow solutions on $n_{\rm H,0}$ (density at $R_{0}=100$ pc), $\alpha$ (where $n_{\rm H} \propto R^{-\alpha}$; eq. (\ref{power law density})), $v_{\rm in}$, and $f_{\rm mix}$. 
The fiducial parameters for these calculations, motivated by local ULIRGs, are: $L_{\rm AGN}=10^{46}$ erg s$^{-1}$, $f_{\rm mix}=1$, $\alpha=1$, $n_{\rm H,0}=100$ cm$^{-3}$, $v_{\rm in}=30,000$ km s$^{-1}$, and $\tau_{\rm in}=1$. 
The figure, in particular, shows how maintaining a shell velocity $v_{\rm s} \approx 1,000$ km s$^{-1}$ at $R_{\rm s} \approx 1$ kpc requires a relatively tenuous ambient medium, $n_{\rm H}(100~{\rm pc})\approx10$ cm$^{-3}$. 
In that case, the momentum boost (quantified by $\dot{M}_{\rm s} v_{\rm s} / (\tau_{\rm in} L_{\rm AGN}/c)$) is $\sim10$, in good agreement with measurements of outflows in AGN-dominated ULIRGs \citep[][]{2010A&A...518L.155F, 2011ApJ...729L..27R, 2011ApJ...733L..16S}. 

In the top left panel (varying $n_{\rm H,0}$), we include examples of the same calculations, but using the inverse Compton cooling time for single-temperature gas (eq. (\ref{Compton time rel})) instead of the approximation for a two-temperature plasma (eq. (\ref{tp cool})). 
Without accounting for the slow down of inverse Compton cooling owing to two-temperature effects, the effective cooling time of shocked wind is underestimated, leading to solutions that depart from energy conservation for high ambient gas densities. 
This effect is potentially very important for outflows that are launched in the nuclei of ULIRGs. 

The examples in the bottom right panel show that mixing is not effective in removing the thermal pressure support of the shocked wind, with a discernible effect on the outflow dynamics only for $f_{\rm mix}\gtrsim10^{3}$, and then only at radii $R_{\rm s}\gtrsim1$ kpc. Although mixing accelerates the cooling rate of the shocked wind via more rapid Coulomb collisions, two-temperature effects nevertheless play a role in keeping the electrons non-relativistic and slowing down inverse Compton cooling (\S \ref{mixing}). In these examples, the large amount of mixing necessary to have a dynamical impact owes principally to the fact that even when the cooling rate limiting step is free free emission, the pre-mixing temperature of the shocked wind is so high ($T_{\rm sh}(v_{\rm in}) \sim 10^{10}$ K) that mixing must reduce it by a large factor before $t_{\rm ff}/t_{\rm cr}<1$. The impact of mixing is also limited at small radii by the constraint that the effective $f_{\rm mix}$ cannot exceed the value $f_{\rm. mix}^{\rm max}(R_{\rm s})$ attained when all the swept-up gas mass is mixed with the shocked wind (\S \ref{mixing}).

The lower left panel of Figure \ref{numsols} includes examples of winds with $v_{\rm in}=1,000$ km s$^{-1}$ and $v_{\rm in}=3,000$ km s$^{-1}$. These slower winds are proxies for outflows accelerated by radiation pressure on dust \citep[e.g.,][]{2012arXiv1204.0063R}. 
While these examples initially experience momentum-conserving and partially radiative bubble phases, they eventually transition to the energy-conserving regime and their momentum flux is also boosted significantly. 
On the other hand, these slower winds quickly slow down to velocities $\ll 1,000$ km s$^{-1}$ as they sweep up the ambient medium, a point to which we return in \S \ref{driving mechanisms}. 

Although it is difficult for an AGN outflow to instantaneously unbind the bulk of the molecular disk (eq. (\ref{outward to gravity force})), it remains an open question whether such outflows will eventually empty the disk of most of its gas, as suggested by the short gas depletion time scales inferred by \cite{2011ApJ...733L..16S}. 
Two limits are envisioned. 
In the first limit, the outflow recently ejected the gas normal to the molecular disk, in which it was originally confined. In an instantaneous sense, the outflow has a large momentum flux and is highly mass-loaded. 
However, the wind subsequently escapes easily along paths of least resistance now cleared and has little global impact on the galaxy. 
In the other limit, the AGN wind is still only effective at blowing out the ambient gas in its immediate vicinity, but will eventually incorporate most of the ISM of the galaxy as gas is driven into the nucleus by gravitational torques. 
Such gravitational torques are expected in local ULIRGs, which were likely triggered by recent galaxy mergers \citep[e.g.,][]{1988ApJ...325...74S, 1994ApJ...431L...9M}. 

We plan to investigate the global impact of AGN-driven outflows in more detail in future work.

\begin{figure}
\begin{center} 
\includegraphics[width=0.48\textwidth]{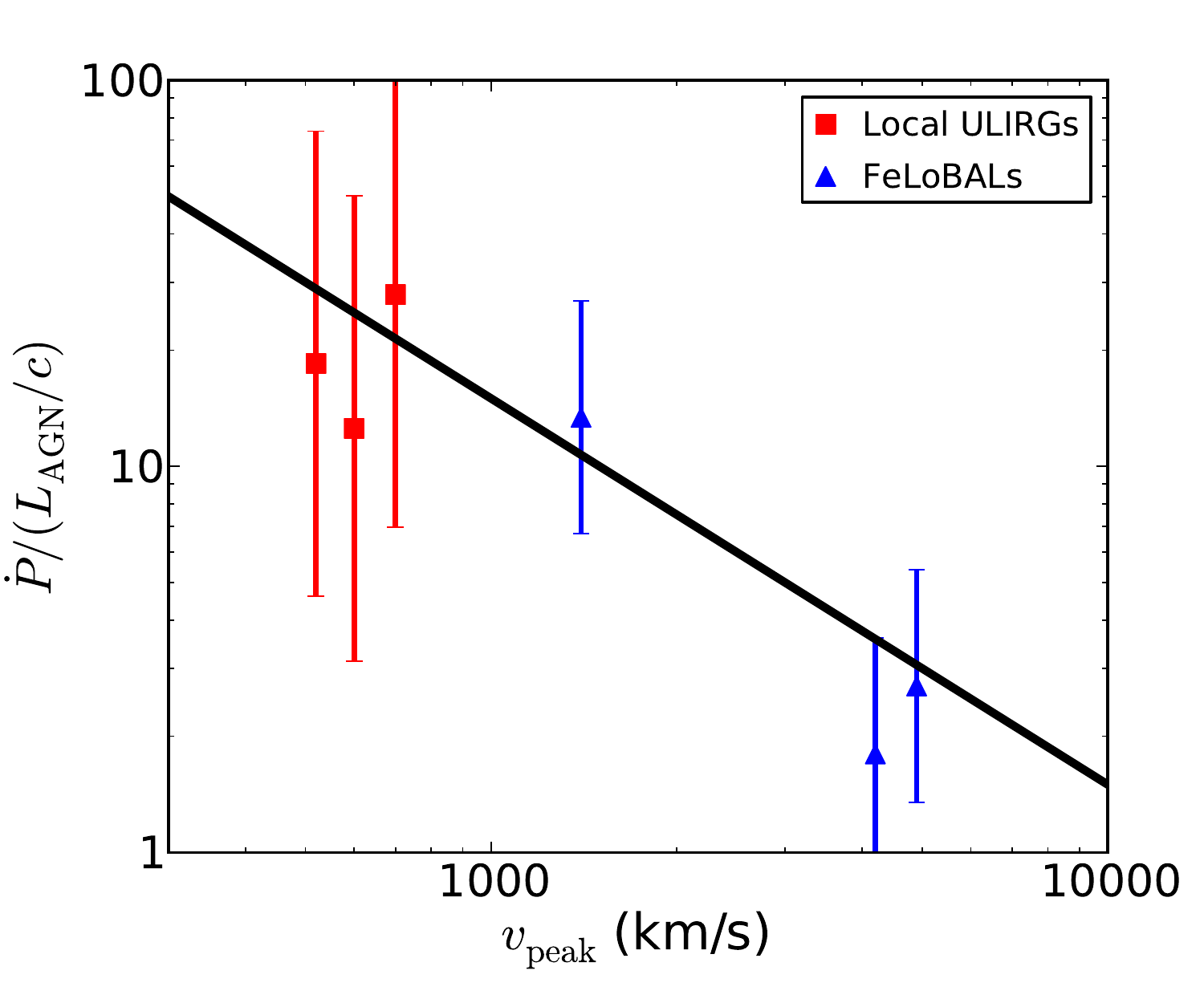}
\end{center}
\caption[]{Momentum flux of the galaxy-scale outflow, relative to $L_{\rm AGN}/c$, as a function of velocity of peak absorption for systems in which the outflow is believed to be driven by the AGN.  
\emph{Solid squares:} Molecular outflows observed in OH by \cite{2011ApJ...733L..16S}, for the ULIRGs with strong AGN contribution. An uncertainty of a factor 4 is assumed in each case, dominated by the uncertainty on the mass outflow rate (the momentum flux is estimated as the mass outflow rate times the velocity of peak absorption). From left to right: IRAS 13120-5453, Mrk 231, and IRAS 08572+3915. 
\emph{Solid triangles:} FeLoBAL quasars interpreted with the radiative shock model of \cite{2011arXiv1108.0413F}, using data from \cite{2009ApJ...706..525M}, \cite{2010ApJ...709..611D}, and \cite{2010ApJ...713...25B}. We assume a factor of 2 uncertainty from the outflow covering factor and the photoionization modeling. From left to right: QSO 2359-1241, SDSS J0318-0600, and SDSS J0838+2955.
The black solid curve shows the prediction $\dot{P} / (L_{\rm AGN}/c) = v_{\rm in} / 2 v_{\rm peak}$ implied by energy conservation, assuming $v_{\rm in}=0.1c$ and $\dot{P}_{\rm in}=L_{\rm AGN}/c$ (eq. \ref{P boost E cons}). 
}
\label{Pdot vs vpeak figure} 
\end{figure}

\subsection{Momentum boost}
\label{momentum boost sec} 
For outflows that are energy-conserving during most of their expansion, the expected momentum boost can be estimated simply using
\begin{equation}
\dot{M}_{\rm s} v_{\rm s}^{2} \approx \frac{1}{2} \dot{M}_{\rm in} v_{\rm in}^{2},
\end{equation} 
again assuming that $\approx 1/2$ of the small-scale wind energy goes into bulk motion of the swept-up gas. 
Defining the large-scale momentum flux as $\dot{P}_{\rm s} \equiv \dot{M}_{\rm s} v_{\rm s}$ and $\dot{P}_{\rm in} \equiv \dot{M}_{\rm in} v_{\rm in}$,
\begin{equation}
\label{P boost E cons}
\frac{\dot{P}_{\rm s}}{\dot{P}_{\rm in}} \approx \frac{1}{2} \frac{v_{\rm in}}{v_{\rm s}}
\end{equation}
\citep[see also][]{2012arXiv1201.0866Z}.  
Figure \ref{Pdot vs vpeak figure} compares this prediction with the observed outflows, as a function of the inferred velocity of peak absorption $v_{\rm peak}$. 
The theoretical curve assumes $v_{\rm in}=0.1c$. 
Interestingly, this prediction of energy conservation is in good agreement with the data (Veilleux \& Rupke 2011, Sturm et al. 2011\nocite{2011ApJ...729L..27R, 2011ApJ...733L..16S}), both in normalization and in slope. 
As the numerical examples in Figure \ref{numsols} show, gravity acts to suppress the momentum flux relative to the simple estimate in equation (\ref{P boost E cons}) when it becomes important as the flow decelerates.  

\section{Discussion}
\label{discussion}

\subsection{Summary}
We have studied the conditions under which galaxy-scale AGN outflows conserve energy versus momentum. 
We focused primarily on winds with initial velocity $v_{\rm in} \gtrsim 10,000$ km s$^{-1}$, as may be launched from the black hole accretion disk, but also considered slower examples as proxies for winds accelerated by radiation pressure on dust. 
Our study was motivated in part by observations of galaxy-scale outflows driven by AGN suggesting that they carry momentum fluxes many times the value $\dot{P} \sim L_{\rm AGN}/c$ expected if the outflows are accelerated by radiation pressure and each photon from the AGN scatters once with the surrounding medium \citep[e.g.,][]{2011ApJ...733L..16S}. 
Numerical simulations of AGN feedback also appear to require $\dot{P} \gg L_{\rm AGN}/c$ in order to reproduce the observed $M_{\rm BH}-\sigma$ relation \citep[][]{2011MNRAS.412.1341D, 2011MNRAS.tmp.2150D}. 
If an outflow conserves energy rather than momentum, then hot shocked gas can do work on the swept-up material, thus boosting the momentum flux, increasing the dynamical importance of the AGN outflow, and potentially providing a simple explanation for the observations of high values of $\dot{P}/(L_{\rm AGN}/c)$. 

We first developed a theory of fast AGN wind bubbles under the assumption of spherical symmetry, building on analogous work in the context of stellar wind bubbles \citep[e.g.,][]{1977ApJ...218..377W, 1992ApJ...388..103K}.  The principal improvement of our analysis over previous work \citep[e.g.,][]{2003ApJ...596L..27K, 2011MNRAS.415L...6K} is a more complete treatment of the cooling of the shocked wind (\S \ref{shocked wind cooling}).  We showed that the time scale for Coulomb coupling between protons and electrons in the shocked wind can be sufficiently long that the post-shock gas develops a two-temperature plasma structure, significantly modifying its effective cooling rate.  The competition between inverse Compton cooling and Coulomb heating of the electrons causes them to remain non-relativistic, and introduces Coulomb coupling as the rate limiting step for removing the thermal pressure support provided by the protons.  Both of these effects slow down the cooling of the protons in the shocked wind relative to simpler estimates that assume that the electrons and the protons have the same temperature (and thus that the electrons are relativistic).  We also carried out our analysis for ambient density profiles of arbitrary slope, rather than focusing on the isothermal case as in most previous studies.  

From our spherically-symmetric analysis, we conclude that fast AGN outflows conserve energy in a significantly broader range of circumstances than previously appreciated (\S \ref{outflow solutions}). For instance, a wind with $v_{\rm in}=30,000$ km s$^{-1}$ and $\dot{P}_{\rm in}=L_{\rm AGN}/c$ ($\tau_{\rm in}=1$) driven by a black hole of mass $M_{\rm BH}=10^{8}$ M$_{\odot}$ radiating at its Eddington limit is predicted to be effectively energy-conserving everywhere for ambient media of uniform density as high as $n_{\rm H}\approx8\times10^{6}$ cm$^{-3}$ (Appendix \ref{self similar analysis}). 
If the outflow is energy-conserving, then the momentum boost of the swept-up material relative to the nuclear wind can reach $\dot{P}_{\rm s} / \dot{P}_{\rm in} \sim v_{\rm in}/2 v_{\rm s}$. 
Thus, an energy-conserving outflow of velocity $v_{\rm s}\sim 1,000$ km s$^{-1}$ driven by a nuclear wind with $v_{\rm in}\sim 30,000$ km s$^{-1}$ would have experienced a momentum flux boost by a factor up to $\sim 15$. 
Interestingly, these estimates are in good agreement with observations of galaxy-scale outflows believed to be driven by AGN in local ULIRGs \citep[see Fig. \ref{Pdot vs vpeak figure};][]{2011ApJ...729L..27R, 2011ApJ...733L..16S}. 

We also considered how three-dimensional effects might affect our conclusion.  Mixing of cold gas with the shocked wind, either from ablation of clumps in the ambient medium or via instabilities at the interface with the swept-up shell, can accelerate the cooling rate.  Depending on the amount of mixing, the rate limiting step for the shocked wind cooling can be either: 1) Coulomb collisions; 2) non-relativistic inverse Compton cooling; or 3) free free cooling.  Although non-relativistic inverse Compton cooling can be somewhat more rapid than the Coulomb-limited regime for outflows with $v_{\rm in} \gg 10,000$ km s$^{-1}$, it is still relatively slow; catastrophic cooling can only occur when the cooling becomes dominated by free free emission.  However, a large amount of mixing can be tolerated before this happens (\S \ref{mixing}).  For the parameters of our fiducial numerical calculations in \S \ref{outflow solutions}, motived by local ULIRG observations, a cold gas mass $\gtrsim 10^{3}$ times that of the shocked wind mass must be mixed so that cooling is accelerated enough to have a discernible dynamical impact.  This is much more than for the analogous problem of stellar wind bubbles \citep[e.g.,][]{1984ApJ...278L.115M} owing to the higher temperature of the shocked wind for the AGN case with $v_{\rm in} \sim 0.1c$, which requires more mixing before the cooling time is reduced to less than $t_{\rm cr}$ or $t_{\rm flow}$.

Propagation of the outflow along paths of least resistance in an anisotropic, multiphase ambient medium (``leakage'') tends to limit the momentum boost that can be achieved. 
However, only a modest amount of confinement is necessary to provide momentum boosts comparable to those inferred in ULIRGs and luminous quasars. 
In fact, spatially-resolved observations of the luminous quasar Mrk 231 show evidence of an AGN-driven outflow with $v_{\rm s}\sim1,000$ km s$^{-1}$ extending $\sim3$ kpc from the nucleus \citep[][]{2011ApJ...729L..27R}. 
The mean density along the path of propagation of this outflow must be much less than the density of the massive molecular disk in the same system, since otherwise the outflow would have decelerated to much lower velocities (see eq. (\ref{momentum constraint})). 
As Mrk 231 is viewed nearly face-on, we are likely seeing the outflow as it leaks out normal to the galactic plane. 
Nevertheless, if the outflow is powered by a fast nuclear wind, it has decelerated sufficiently to have experienced a momentum boost comparable to the inferred one. 

\subsection{Importance of physical effects and comparison with previous work}
\label{which effects}
In this work we have studied several key physical effects previously neglected in models of the interaction between AGN winds and gas in their host galaxy.   We now summarize each of these important effects in turn.  

The two-temperature cooling effects slow down inverse Compton cooling for $v_{\rm in} \gtrsim10,000$ km s$^{-1}$ (\S \ref{cooling processes}). 
Relative to the assumption that electrons and protons share the same temperature at all times, the cooling time of the shocked wind can be longer by a factor $\gtrsim 10-100$ (Fig. \ref{TpTe figure}). 
Absent substantial mixing of the shocked wind with cooler gas, inverse Compton scattering off the AGN radiation field (with proton cooling limited by Coulomb collisions with electrons) is the dominant cooling process. 
The two-temperature effects thus directly affect the conditions for energy versus momentum conservation. 
In particular, nuclear winds with $v_{\rm in} \sim 0.1c$ can be in the energy-conserving limit even for the high ambient gas densities and intense radiation fields characteristic of the centers of ULIRGs. 
The longer effective cooling time of the protons owing to two-temperature effects is robust to uncertainties in collisionless processes that may heat the electrons faster than Coulomb collisions near the shock.

Previous analytic studies have focused primarily on isothermal ambient gas density profiles ($\alpha=2$).  We have considered more general profiles with $\alpha < 2$ and have shown that the isothermal profile is in fact a singular case. 
This is because it does not have a finite free expansion radius $R_{\rm f}$, so that outflow solutions retain memory of the initial conditions on large scales. 
With an isothermal sphere density profile and a constant gas mass fraction $f_{\rm g}$, momentum conservation implies a constant swept-up shell velocity
\begin{align}
\label{vm king}
v_{\rm m} & = \left( \frac{\tau_{\rm in} G L_{\rm AGN}}{2 f_{\rm g} \sigma^{2} c} \right)^{1/2} \\ \notag
& \approx 130~{\rm km~s^{-1}}~\tau_{\rm in}^{1/2} \left( \frac{f_{\rm g}}{0.17} \right)^{-1/2} \left( \frac{L_{\rm AGN}}{\rm 10^{46}~erg~s^{-1}} \right)^{1/2} \\ \notag 
& ~~~~~~~~~~~~~~~~~~~~~~~~~~~~~~~~~~~~~~~~\times \left( \frac{\sigma}{\rm 200~km~s^{-1}} \right)^{-1}
\end{align}
\citep[][]{2003ApJ...596L..27K}. 
Because $R_{\rm f} \to \infty$, such solutions are not consistent with the assumption $v_{\rm in} \gg v_{\rm m}$.\footnote{Self-consistent physical solutions do exist, but they differ from the self-similar expression in equation (\ref{vm king}). In particular, they are sensitive to the radius at which the wind is launched and start with $v_{\rm s} = v_{\rm in}$} 
The small velocity $v_{\rm s} = v_{\rm m}$ implies a longer $t_{\rm flow} = R_{\rm s} / v_{\rm s}$. Thus, independent of two-temperature cooling effects, we find that cooling is less important (than, e.g., King 2003\nocite{2003ApJ...596L..27K}) when the non self-similarity of realistic models is included. 
The numerical models in Figure \ref{numsols} in fact show that energy-conserving solutions are obtained for parameters that reproduce observed galaxy-scale outflows even when neglecting two-temperature effects. 

In a real galaxy, the cold gas is flattened into a disk and the ISM is highly inhomogeneous. 
Thus, there is no clear reason to favor $\alpha=2$ as representative of the paths of least resistance along which galactic winds effectively propagate, even if the rotation curve is flat (\S \ref{leakage}). 
For example, \cite{2012ApJ...748...36B} infer the pre-outflow gas density profile around the tidal disruption event Swift J164449.3+573451 (assumed to originate from the central black hole) to have a slope $\alpha \approx 1.5$ within 0.5 pc, with significant flattening farther out. 
Our general self-similar analysis in Appendix \ref{self similar analysis} shows that the transition from momentum to energy conservation, and vice versa, is sensitive to $\alpha$. 

Mixing of the shocked wind with cooler gas tends to increase its cooling rate. 
However, the numerical examples in Figure \ref{numsols} show that our conclusions are robust to a large amount of mixing (see also \S \ref{mixing}).

The $M_{\rm BH}-\sigma$ relation \citep[][]{2000ApJ...539L...9F, 2000ApJ...539L..13G, 2002ApJ...574..740T} provides a constraint on whether AGN winds conserve energy or momentum. \cite{2003ApJ...596L..27K} showed that the observed relation can be reproduced in a simple spherically-symmetric model of momentum-conserving outflows.\footnote{\cite{2005ApJ...618..569M} also showed that the observed $M_{\rm BH}-\sigma$ can be derived from a simple Eddington limit argument for radiation pressure on dust. That argument however does not rely on the existence of a wind within the dust sublimation radius.} In an analogous model assuming energy conservation, the predicted black hole mass lies below the $M_{\rm BH}-\sigma$ relation by more than one order of magnitude. 
Because in real galaxies the gas inflows that fuel central black holes are likely to occur through disks, isotropic black hole winds transfer only a fraction of their outward momentum flux to the inflowing gas. 
Thus, winds that are purely momentum-conserving may actually be insufficient to regulate black hole growth at the level implied by the $M_{\rm BH}-\sigma$ relation. 
The estimate in equation (\ref{outward to gravity force}) indicates that a momentum boost $\tau \sim 10-15$, such as can be attained in an energy-conserving outflow, may be necessary to stop inflows from the molecular disks in ULIRGs to the central black holes. 
Comparable momentum flux enhancements are in fact needed to reproduce the $M_{\rm BH}-\sigma$ relation in three-dimensional simulations \cite[e.g.,][]{2011MNRAS.412.1341D}. 

Finally, we note that although our analysis implies that winds driven by AGN are energy-conserving in a substantially wider range of conditions than found by King and collaborators, those authors also predicted a transition to energy conservation on scales $\gtrsim1$ kpc relevant for galactic winds \citep[][]{2011MNRAS.415L...6K}. 
In addition, significant momentum-conserving phases are not ruled out by our arguments. 
If strong observational evidence for momentum conservation is found, our study of the conditions under which it is realized would provide useful constraints on the physical conditions in galactic nuclei.

\subsection{Driving mechanisms}
\label{driving mechanisms}
Our models do not depend on how the wind is launched on small scales, but we did focus on the case of high initial velocities, $v_{\rm in}\gtrsim 10,000$ km s$^{-1}$. 
Furthermore, in order to explain galaxy-scale momentum fluxes $\dot{P} \gtrsim 10 L_{\rm AGN}/c$ inferred from observations by hot gas confinement, we require that the wind starts with $\dot{P}_{\rm in} \sim L_{\rm AGN}/c$. 
We now discuss the evidence that such winds are realized and the physical mechanisms that can drive them. 

UV BALs are observed in up to $40\%$ of quasars \citep[e.g.,][]{2008ApJ...672..108D} and could be present in the majority of quasars if this fraction represents a viewing angle effect \citep[][]{1991ApJ...373...23W}.  
The physical properties of the absorbing gas are generally not well known. 
Photoionization modeling of a subset of low-ionization absorbers indicates that the kinetic luminosity of the outflowing gas can reach a few percent of $L_{\rm AGN}$ \citep[][]{2009ApJ...706..525M, 2010ApJ...709..611D, 2010ApJ...713...25B}.  
While the absorption in those cases likely arises from the interaction of the quasar outflow with the ISM of the galaxy \citep[][]{2011arXiv1108.0413F}, rather than in the immediate vicinity of the black hole, the energetics do suggest that AGN can drive dynamically important winds. 
The electron scattering optical depth around black holes accreting near the Eddington limit is furthermore naturally $\tau_{\rm es} \sim 1$ \cite[e.g.,][]{2003MNRAS.345..657K}. 
Since the rate of gas inflow from galaxy to nuclear scales is not tuned to the Eddington limit, $\tau_{\rm es}$ may in fact exceed unity by a significant factor in many cases. 
X-ray observations confirm that this is the case in the prototypical example Mrk 231 \citep[][]{2004A&A...420...79B}. 
If the nuclear wind is accelerated by radiation pressure on electrons, its momentum flux on small scales is thus $\dot{P}_{\rm in} \gtrsim L_{\rm AGN}/c$ in these circumstances. 
The assumptions of our analysis may thus be generically realized in luminous quasars. 

Dust in the central regions of ULIRGs is also generally optically thick to the UV photons that dominate the radiative output of AGN, as well as to reprocessed infrared radiation \citep[e.g.,][]{2003JKAS...36..167S, 2005ApJ...630..167T}. 
Like the electron scattering opacity, dust opacity is continuous in wavelength and can thus scatter the bulk of the AGN luminosity. 
Furthermore, the large optical depths in ULIRGs imply that the momentum flux imparted on the surrounding medium can potentially be several times $L_{\rm AGN}/c$ owing to photon trapping (Roth et al. 2012; but see Krumholz \& Thompson 2012, Novak et al. 2012)\nocite{2005ApJ...618..569M, 2012arXiv1203.2926K, 2012arXiv1203.6062N}. 
Thus, realistic simulations of AGN feedback in galaxies should include radiation pressure on dust in addition to the fast nuclear winds analyzed in this work. 

On the other hand, radiation pressure on dust cannot readily explain AGN winds commonly observed in UV and X-ray absorption with velocities $\gg 1,000$ km s$^{-1}$. 
In fact, this mechanism operates most effectively near the dust sublimation radius,
\begin{align}
R_{\rm dust} & = \left( \frac{L_{\rm AGN}}{4 \pi \sigma_{\rm SB} T_{\rm sub}^{4}} \right)^{1/2} \\ \notag
& \approx 1~{\rm pc}~\left( \frac{L_{\rm AGN}}{\rm 10^{46}~erg~s^{-1}} \right)^{1/2} \left( \frac{T_{\rm sub}}{\rm 1,200~K} \right)^{-2},
\end{align}
where $\sigma_{\rm SB}$ is the Stefan-Boltzmann constant and $T_{\rm sub}$ is the dust sublimation temperature. 
For massive black holes, this is within the sphere of influence, where the local escape speed is
\begin{equation}
v_{\rm e}(R) \approx {\rm 930~km~s^{-1}} \left( \frac{M_{\rm BH}}{\rm 10^{8}~M_{\odot}} \right)^{1/2} \left( \frac{R}{\rm 1~pc} \right)^{-1/2}.
\end{equation}
Radiative transfer calculations show that radiation pressure on dust in AGN can accelerate winds to comparable velocities, but not much more \citep[][]{2012arXiv1204.0063R}. 
Since outflows also decelerate substantially as they sweep up ambient medium (unless the density profile is very steep), winds driven by radiation pressure on dust are therefore limited to a few $1,000$ km s$^{-1}$ at most.

In addition to radiation pressure on electrons and on dust, radiation pressure on UV lines \citep[e.g.,][]{1995ApJ...451..498M, 2000ApJ...543..686P}, hydromagnetic processes \citep[e.g.,][]{1992ApJ...385..460E, 1994ApJ...434..446K}, and Compton heating \citep[e.g.,][]{2005MNRAS.358..168S} may contribute to accelerating outflows. 
An important goal for future work is to determine the exact physical mechanisms through which a central black hole couples to the surrounding galaxy.

\subsection{Observational signatures of shocked AGN wind bubbles}
\label{observational signatures}
Energy-conserving outflows predict the existence of shocked wind bubbles. 
We argue here that such bubbles are consistent with current observations and discuss ways of probing them in more detail in the future. 

\subsubsection{Direct wind bubble emission}
The free free, inverse Compton, and synchrotron emission from the \emph{thermal} electrons in the shocked wind bubble are estimated in Appendix \ref{shocked wind emission}. 
The very hard X-ray ($k T_{\rm e} \approx 260~{\rm keV}~(T_{\rm e}/3\times10^{9}~{\rm K})$) and relatively faint thermal free free will be extremely challenging to detect directly even with new-generation observatories such as \emph{NuSTAR}.\footnote{http://www.nustar.caltech.edu/} 
Furthermore, synchrotron emission from the thermal electrons has a characteristic frequency $\nu_{\rm c} \sim $ 4 Hz $\gamma^2$ $(B/{\rm 1 \mu G})$, which is un-observable due to the plasma frequency $\sim 3$ MHz of the Galactic warm interstellar medium \citep[][]{2010MNRAS.406..863L}.

In terms of both luminosity and energy band, the most promising signature of the thermal electrons is the inverse Compton continuum that they should produce at $\sim k T_{\rm X} \sim$ keV when they interact with the AGN radiation field \citep[see also][]{2010MNRAS.402.1516K}. At best, when the inverse Compton cooling time is short, this signal will be no more than a few percent of the radiative AGN luminosity (eq. \ref{mechanical fraction}). 
Indeed, if $\dot{M}_{\rm in} v_{\rm in} = L_{\rm AGN}/c$, then a fraction
\begin{equation}
\label{mechanical fraction}
\frac{(\dot{M}_{\rm in} v_{\rm in}^{2} / 2)}{L_{\rm AGN}} = \frac{v_{\rm in}}{2 c} = 0.05 \left( \frac{v_{\rm in}}{\rm 30,000~km~s^{-1}} \right)
\end{equation}
of the radiative luminosity is converted into wind mechanical energy that can then be radiated away.
Nevertheless, this may be detectable. 

\cite{2011MNRAS.413.1251P} report a possible detection of the inverse Compton continuum in the Seyfert 1 galaxy NGC 4051, with a luminosity comparable to the mechanical luminosity of the wind in that system. 
Although those authors interpret this as evidence of shocked wind cooling and thus of a momentum-conserving outflow, this does not necessarily follow. 
In fact, the inverse Compton emission quantifies the energy drain from the shocked electrons, \emph{but it is the protons that provide the majority of the thermal pressure of the shocked wind} (\S \ref{shocked wind cooling}). 
If the electrons experience a significant amount of collisionless heating at the shock, but are predominantly coupled to the protons via Coulomb collisions over the longer time scale relevant for shocked wind cooling, then inverse Compton emission with a luminosity similar to the wind mechanical luminosity could result without removing the thermal pressure provided by the protons.

Relativistic particles accelerated either at the forward or reverse shock \citep[e.g.,][]{1978ApJ...221L..29B} may also produce non-thermal radio and $\gamma-$ray emission. 
While such emission does not provide direct information on the state of the gas inside the shocked wind bubble, it probes the mechanical energy processed through the shocks and can contribute significantly to the spectral energy distribution of systems hosting AGN. 
Kiloparsec-scale radio synchrotron is commonly observed from AGN hosts \citep[e.g.,][]{1993ApJ...419..553B, 2006AJ....132..546G} and usually interpreted as being powered by either a starburst or by a black hole jet interacting with the ISM. 
Cosmic ray electrons accelerated by a wide-angle AGN wind may provide an alternative explanation in some cases. 
In particular, Mrk 231 has spatially extended radio emission on scales of $\sim 100$ pc to $\sim 25$ kpc \citep[][]{1999ApJ...516..127U} which the AGN wind can easily account for energetically (eq. (\ref{nonthermal synch})). 

\subsubsection{Effects of wind bubbles on surroundings}
If the shocked protons do not cool and the outflow is energy-conserving, then the hot gas will tend to expand along paths of least resistance (\S \ref{leakage}). 
For an AGN outflow launched at the center of a galactic disk, the simplest outcome would be bipolar bubbles expanding above and below the disk plane. 

\cite{2012ApJ...746...86G} report the discovery of such bubbles around the luminous obscured quasar SDSS J1356$+$1026 at $z=0.123$. 
The bubbles extend $\sim 10$ kpc on each side of the galaxy, with a probable de-projected expansion velocity $\sim 1,000$ km s$^{-1}$. 
These bubbles, most likely inflated by the quasar, are strongly suggestive of buoyant hot gas, such as a shocked wind that has not cooled. 
In this case, the observed [OIII] and H$\beta$ emission is consistent with being powered by photoionization by an AGN spectrum and thus likely traces the swept-up up ambient medium. 
We must note, though, that SDSS J1356$+$1026 is the only one out of 15 obscured quasar studied by \cite{2011ApJ...732....9G} showing such prominent bubbles. 
Although these bubbles are therefore not ubiquitous in an instantaneous sense, simulations of quasars triggered in galaxy mergers indicate that the accreting black hole may spend up to $\sim 90\%$ of its lifetime in a buried phase, during which the hot gas may not leak effectively. 
It is thus possible that most of the objects observed by \cite{2011ApJ...732....9G} will eventually inflate bubbles similar to SDSS J1356$+$1026.

\emph{Fermi}\footnote{http://fermi.gsfc.nasa.gov/} also recently detected bubbles of a similar scale in our Galaxy, clearly originating from the Galactic Center \citep[][]{2010ApJ...717..825D, 2010ApJ...724.1044S}. 
In our notation, the bubbles are inferred to have an age $t_{\rm flow} = R_{\rm s} / v_{\rm s} \sim 10^{7}~{\rm yr}~(R_{\rm s}/{\rm 10~kpc})(v_{\rm s}/{\rm 1,000~km~s^{-1}})^{-1}$. 
Interestingly, observations also reveal that a starburst in the central parsec occurred $6 \pm 2$ Myr ago \citep[][]{2006ApJ...643.1011P}, perhaps accompanied by an accretion event onto Sgr A$^{\star}$ \citep[][]{2003ApJ...590L..33L}. 
Suppose that the black hole accretion event lasted $\lesssim 1$ Myr and inflated a shocked wind bubble with $T_{\rm sw}\sim10^{9}$ K ($v_{\rm in} \sim 10,000$ km s$^{-1}$). 
The bubble may have since expanded by a factor $\sim 10$ in linear dimension, or $\sim 10^{3}$ in volume. 
Adiabatic cooling ($T \propto \rho^{2/3}$) would then imply that the bubble would now have an interior temperature $\sim 10^{7}$ K. 
Thermal free free emission from gas at this temperature is consistent with 1.5 keV soft X-ray features coincident with the \emph{Fermi} bubbles previously detected by \emph{ROSAT}\footnote{http://heasarc.gsfc.nasa.gov/docs/rosat/rosat.html} \citep[][]{2003ApJ...582..246B}. 
The bipolar bubble morphology would then naturally result from confinement by the massive molecular disk in the inner Galaxy \citep[e.g.,][]{2012arXiv1203.3060Z}. 
Thus, the \emph{Fermi} bubbles may be the relics of an energy-conserving AGN outflow. 

\subsection{Conclusion and future directions}
Many questions remain open regarding the global impact of AGN winds on galaxy evolution. 
In particular, it is not clear if (and if so, how) such winds have a major effect on star formation in galaxies. 
While a quasar outburst can easily release enough energy to completely unbind the interstellar medium of a galaxy \citep[e.g.,][]{1998A&A...331L...1S}, it is not trivial to couple this energy effectively to a galactic disk. 

Indeed, galactic disks subtend only a small fraction of the solid angle around the black hole and nuclear outflows are likely to preferentially escape normal to the disk. 
Although OH observations in local ULIRGs suggest short gas depletion time scales \citep[][]{2011ApJ...733L..16S}, most of the CO flux in some of the same systems is observed to originate in a narrow component tracing the star-forming molecular disk \citep[e.g.,][]{2010A&A...518L.155F}, rather than a fast outflow. 
It may be possible to reconcile these observations if the gas from the disk is rapidly funneled into the nucleus by strong gravitational torques and only gradually incorporated into the wind. 
Alternatively, the wind may be gradually mass loaded by efficient mixing at the wind-disk interface. 
It is also possible that the large mass outflow rates have short duty cycles, so that they do not necessarily remove a large mass fraction. 
As mentioned above, \cite{2011ApJ...732....9G} observe significant, non-gravitational disturbances over the entire host galaxies of luminous obscured quasars at $z < 0.5$ and also find evidence that their interstellar media are completely ionized by the quasar. On the other hand, the case for effective gas ejection and star formation truncation in these systems is less clear.  

The galaxy-scale impact of AGN winds depends on three-dimensional effects, non-linear interactions between stellar and black hole feedback processes \citep[e.g.,][]{2012arXiv1203.3802B}, and the detailed structure of the ISM. It will thus be fruitful to incorporate the physical insights gained in this work into more detailed numerical simulations. In Appendix \ref{numerical simulations}, we outline how our results can be used to implement the effects of fast nuclear winds more realistically in such simulations. 

\section*{Acknowledgments}
We thank Phil Hopkins, Andrew King, Chris McKee, Norm Murray, and Nathan Roth for useful discussions. 
We are also grateful to the referee, Joop Schaye, for a detailed and constructive report. 
CAFG is supported by a fellowship from the Miller Institute for Basic Research in Science and NASA grant 10-ATP10-0187.  EQ is supported in part by the David and Lucile Packard Foundation and the Thomas and Alison Schneider Chair in Physics at UC Berkeley.

\appendix


\section{Energy and momentum conservation conditions for self-similar outflows}
\label{self similar analysis}
In \S \ref{outflow solutions}, we presented examples of numerical solutions relevant for observed galaxy-scale AGN outflows and concluded that they were energy-conserving. Here we present an analytic analysis of self-similar solutions to determine the dependences on problem parameters and show that energy conservation is realized over a wide parameter space. 

To ensure self-similarity, we assume a power-law ambient density profile (eq. (\ref{power law density})) and neglect gravity. 
We restrict ourselves to decelerating solutions with finite free expansion radius (see below), corresponding to $\alpha<2$. 
We focus on the two-temperature plasma cooling case, which should be realized for fast nuclear winds in spherical symmetry, and on solutions that begin in the energy or momentum-conserving limits. 
We then quantify the transition to a partially radiative bubble in each case. 
At this point, the solution deviates from self-similarity and becomes neither fully energy or momentum-conserving. 

\begin{figure}
\begin{center}
\includegraphics[width=0.47\textwidth]{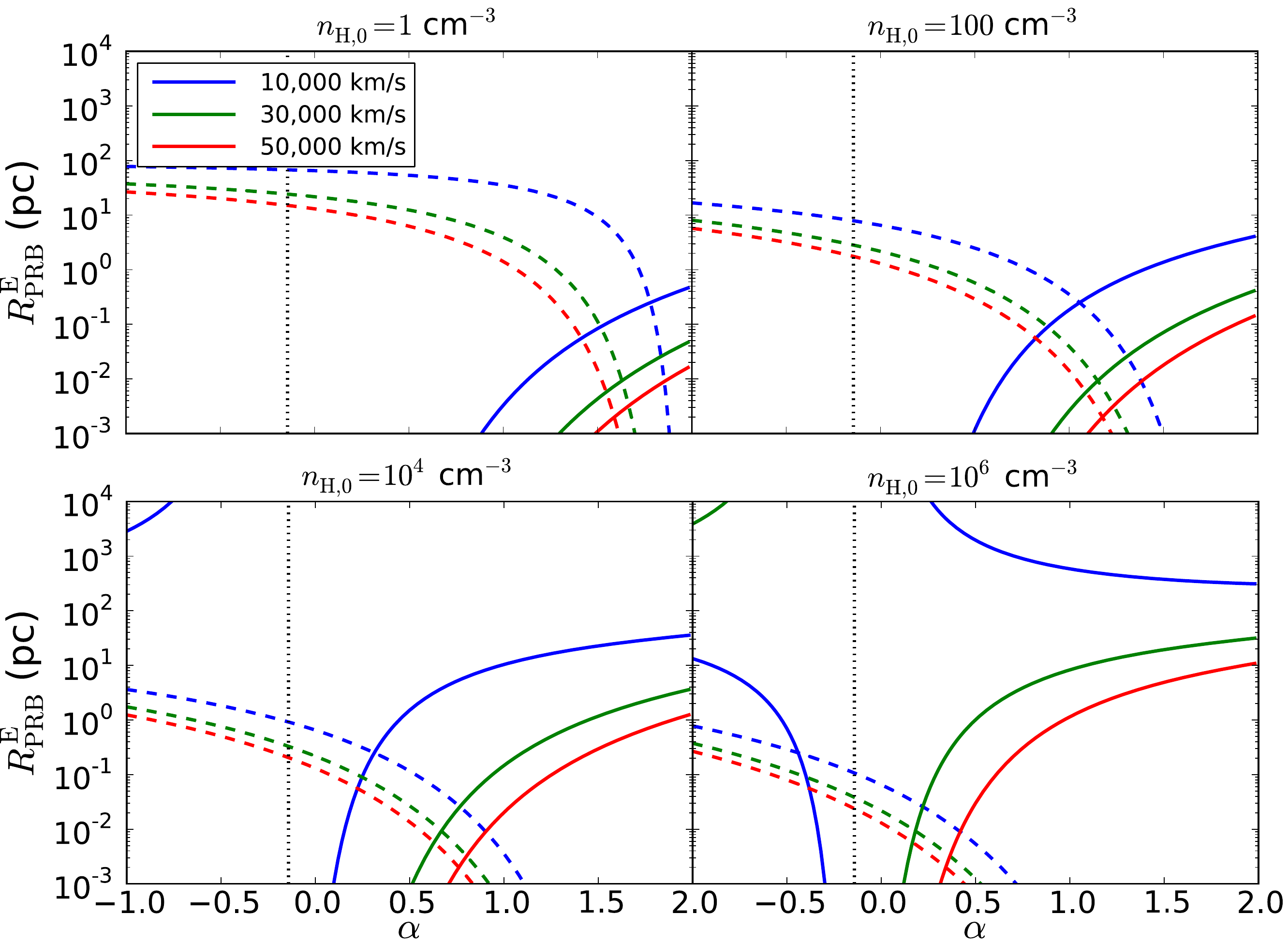}
\end{center}
\caption[]{Radius at which self-similar energy-conserving solutions transition to a partially radiative bubble as a function of the slope of the ambient density profile (solid curves). Fiducial AGN parameters ($L_{\rm AGN}=10^{46}$ erg s$^{-1}$, $\tau_{\rm in}=1$) are assumed, and $v_{\rm in}=10,000,~30,000,~50,000$ km s$^{-1}$ is varied. The different panels correspond to different densities at the reference radius $R_{0}=100$ pc. The free expansion radius, $R_{\rm f}$, is shown by a dashed curve in each case. The vertical dotted line at $\alpha=-1/7$ separates cases where the solution can be energy conserving everywhere (when $R_{\rm PRB}^{\rm E}<R_{\rm f}$ for $-1/7<\alpha<2$) and the cases where solutions transition to a partially radiative bubble at a finite radius ($\alpha<-1/7$).
}
\label{Rc fig} 
\end{figure}

\begin{figure}
\begin{center}
\includegraphics[width=0.47\textwidth]{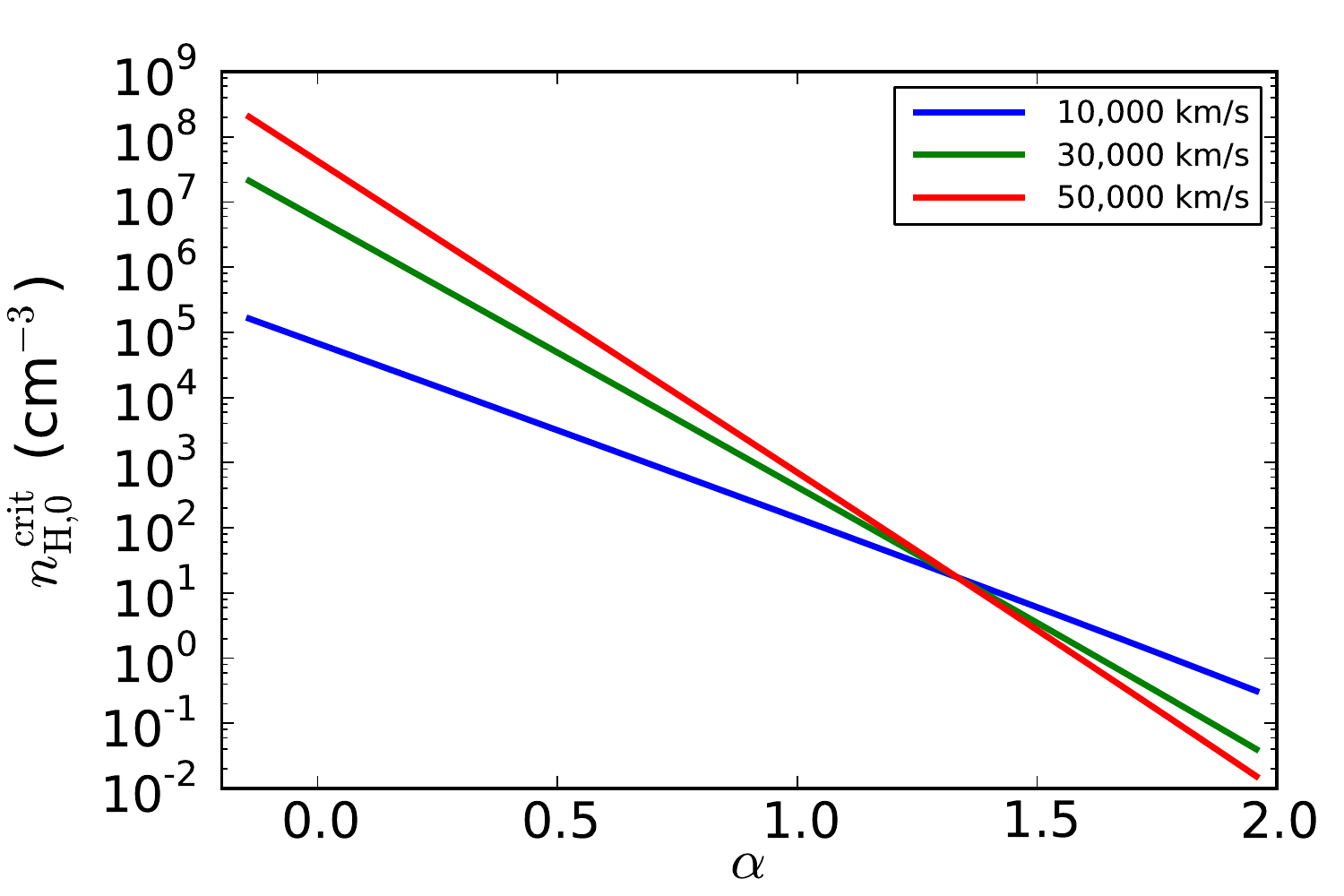}
\end{center}
\caption[]{Maximum density of the ambient medium at $R_{0}=100$ pc for which the outflow solution is purely energy-conserving ($R_{\rm PRB}^{\rm E}<R_{\rm f}$) versus the slope of the ambient medium profile. Purely energy-conserving outflows are only possible for $-1/7 < \alpha < 2$. Fiducial AGN parameters ($L_{\rm AGN}=10^{46}$ erg s$^{-1}$, $\tau_{\rm in}=1$) are assumed.
}
\label{rhoc fig} 
\end{figure}

\subsection{Solutions that begin energy-conserving}
\label{energy conservation conditions} 
Consider first solutions that begin in the energy-conserving limit. Again assuming that 1/2 of the mechanical energy injected by the AGN goes into kinetic motion of the swept-up gas, energy conservation implies
\begin{equation}
\label{E cons eq}
L_{\rm in} t = M_{\rm s} v_{\rm s}^{2},
\end{equation}
where
\begin{align}
M_{\rm s} & = 4 \pi \int_{0}^{R_{\rm s}} dR R^{2} \rho_{\rm g}(R) \\ \notag
& = \frac{4 \pi \rho_{0} R_{0}^{\alpha}}{3-\alpha} R_{\rm s}^{3-\alpha}.
\end{align}
For a self-similar solution of the form $R_{\rm s}= A_{\rm E} t^{\beta_{\rm E}}$ ($\Rightarrow v_{\rm s}=\dot{R}_{\rm s}=\beta_{\rm E} R_{\rm s} / t$),  equation (\ref{E cons eq}) implies
\begin{equation}
A_{\rm E} = \left[ \frac{(5-\alpha)^{2} (3-\alpha) L_{\rm in}}{36 \pi \rho_{0} R_{0}^{\alpha}} \right]^{\frac{1}{5-\alpha}}
\end{equation}
and
\begin{equation}
\beta_{\rm E} = \frac{3}{5 - \alpha}. 
\end{equation}
We can then explicitly relate $v_{\rm s}$ and $R_{\rm s}$:
\begin{equation}
v_{\rm s} = \frac{3}{2}
\left[
\frac{(3-\alpha) L_{\rm AGN} \tau_{\rm in} v_{\rm in}}{\pi (5 - \alpha) c R_{0}^{\alpha} \rho_{0}}
\right]^{1/3} 
R_{\rm s}^{(\alpha-2)/3}~~~~~(E~{\rm cons.}).
\end{equation}
The free expansion radius is defined as the solution to $\dot{M}_{\rm in} t = M_{\rm s}(R_{\rm s}(t))$:
\begin{equation}
\label{Rf gen}
R_{\rm f} = 
\left[
\frac{2^{3/2} (5 - \alpha) \pi c R_{0}^{\alpha} \rho_{0} v_{\rm in}^{2}}{3 (3 - \alpha) \tau_{\rm in} L_{\rm AGN}}
\right]^{1/(\alpha - 2)}.
\end{equation}
Inside $R_{\rm f}$, the integrated mass from the nuclear wind exceeds that of the swept-up ambient medium and cooling losses are dynamically unimportant. 

In the energy-conserving limit,
\begin{align}
\label{t ratio gen}
\frac{t_{\rm p,cool}}{t_{\rm flow}} &= \frac{9 \times 3^{1/15} \pi^{1/3}}{32\times 2^{1/5}} \\ \notag 
&~~~~~~~~~~ \times \left[
\left( \frac{\alpha - 3}{\alpha - 5} \right)^{7}
\frac{m_{\rm e}^{6} m_{\rm p}^{21} \mu^{9} c^{17} \tau_{\rm in} v_{\rm in}^{31}}{e^{24} \sigma_{\rm T}^{9} (\ln{\Lambda})^{6} \rho_{0}^{7} R_{0}^{7\alpha} L_{\rm AGN}^{8}}
\right]^{1/15} \\ \notag
&~~~~~~~~~~~~~~~~~~~~ \times R_{\rm s}^{(1+7\alpha)/15}.
\end{align}
The transition to a partially radiative bubble occurs when this ratio is unity, i.e. at 
\begin{equation}
\label{Rc gen}
R_{\rm PRB}^{\rm E} = \left[
\frac{2^{78}}{3^{31} \pi^{5}}
\left(
\frac{\alpha - 5}{\alpha - 3}
\right)^{7} 
\frac{e^{24} \sigma_{\rm T}^{9} (\ln{\Lambda})^{6} \rho_{0}^{7} R_{0}^{7 \alpha} L_{\rm AGN}^{8}}{c^{17} m_{\rm e}^{6} m_{\rm p}^{21} \mu^{9} \tau_{\rm in} v_{\rm in}^{31}}
\right]^{1/(1+7\alpha)}.
\end{equation}
Figure \ref{Rc fig} compares $R_{\rm PRB}^{\rm E}$ and $R_{\rm f}$ for representative parameters. 

If and only if $-1/7<\alpha < 2$, $t_{\rm p,cool}/t_{\rm flow}$ increases with $R_{\rm s}$ and solutions can be energy-conserving at all radii past $R_{\rm f}$. 
Here, the condition $\alpha<2$ simply re-iterates the assumption made earlier to ensure that the wind has a small free expansion radius. 
For fixed AGN parameters and ambient profile slope, we define the critical density $\rho_{\rm 0,c}$ such that $R_{\rm PRB}^{\rm E}/R_{\rm f}=1$. This is the maximum ambient gas density at $R_{0}$ such that the solution is effectively energy-conserving everywhere:
\begin{align}
\rho_{\rm 0}^{\rm crit} &=  2^{3(3\alpha -7)/2} 3^{(21 - 8\alpha)/5} \pi^{(3-4\alpha)/5} c^{(11-8\alpha)/5} \left( \frac{\alpha - 3}{\alpha - 5} \right) \\ \notag 
& \times \left[ \frac{e^{8} (\ln{\Lambda})^{2} \sigma_{\rm T}^{3}}{m_{\rm e}^{2} m_{\rm p}^{7} \mu^{3}} \right]^{(\alpha - 2)/5}
\frac{L_{\rm AGN}^{\alpha - 1} \tau_{\rm in}^{(1 + 2\alpha)/5} v_{\rm in}^{4-3\alpha}}{R_{0}^{\alpha}}.
\end{align}
Figure \ref{rhoc fig} shows $n_{\rm H,0}^{\rm crit}\equiv \rho_{\rm 0}^{\rm crit}/m_{\rm p}$ as a function of $\alpha$ for $R_{0}=100$ pc, fiducial AGN parameters ($L_{\rm AGN}=10^{46}$ erg s$^{-1}$, $\tau_{\rm in}=1$), and $v_{\rm in}=10,000,~30,000$ and 50,000 km s$^{-1}$. 
For example, for a wind with $v_{\rm in}=30,000$ km s$^{-1}$ blown into an ambient medium of uniform density, the outflow is energy-conserving at all times for $n_{\rm H}<8\times10^{6}$ cm$^{-3}$. 

For $\alpha<-1/7$, solutions can begin energy-conserving at $R_{\rm f}$ but transition to a partially radiative bubble at $R_{\rm PRB}^{\rm E}>R_{\rm f}$. However, the conditions for such a transition to occur are quite extreme; the large ambient densities required would imply much more severe deceleration of the swept-up ambient medium than is observed in the systems of interest (see the momentum constraint in eq. (\ref{momentum constraint})).

\begin{figure}
\begin{center}
\includegraphics[width=0.47\textwidth]{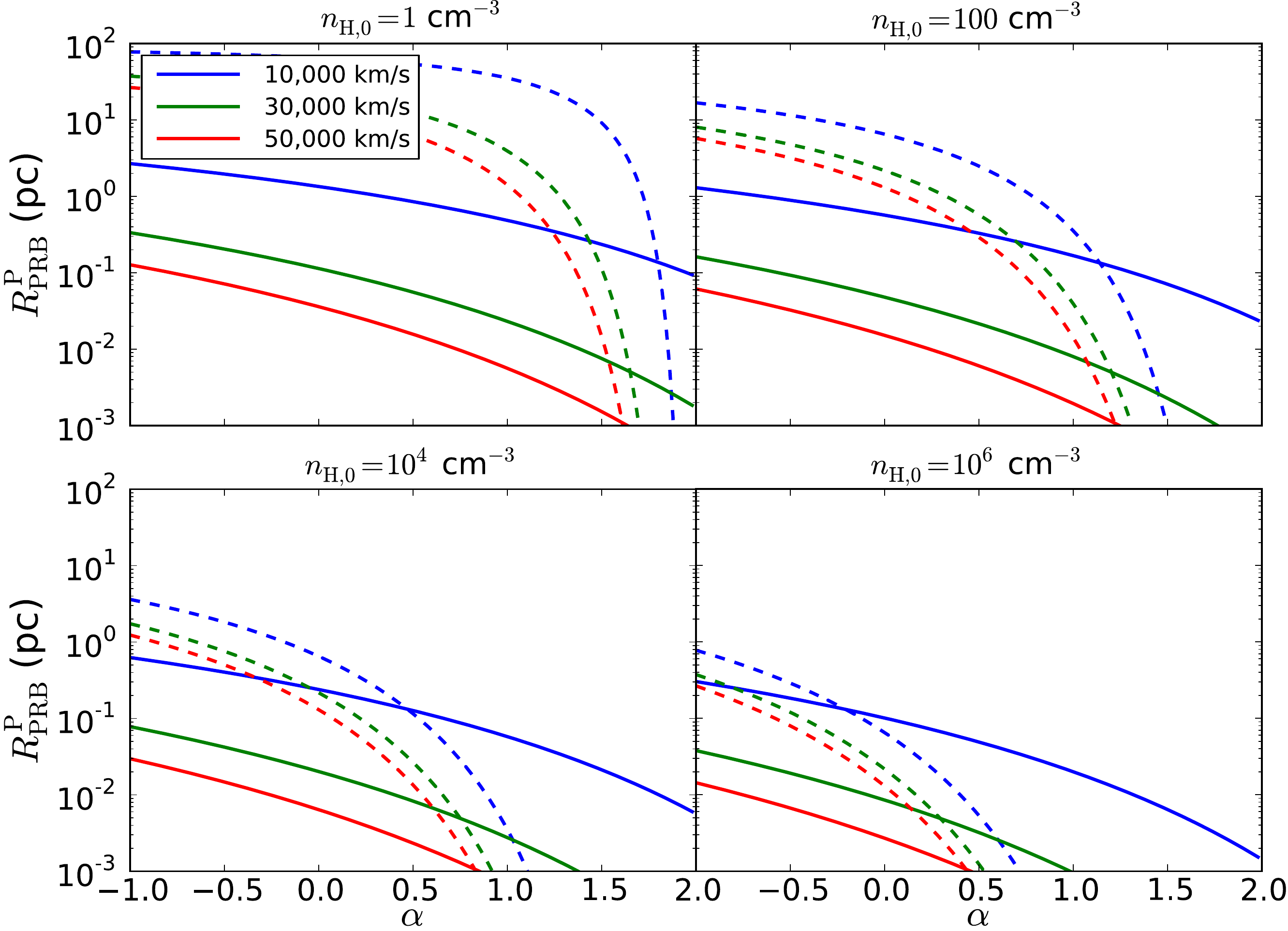}
\end{center}
\caption[]{Radius at which self-similar momentum-conserving solutions transition to a partially radiative bubble as a function of the slope of the ambient density profile (solid curves). Fiducial AGN parameters ($L_{\rm AGN}=10^{46}$ erg s$^{-1}$, $\tau_{\rm in}=1$) are assumed. The different panels correspond to different densities at the reference radius $R_{0}=100$ pc. The free expansion radius, $R_{\rm f}$, is shown by a dashed curve in each case.
}
\label{Rcp fig} 
\end{figure}

\subsection{Solutions that begin momentum-conserving}
\label{momentum to energy transition} 
In a manner analogous to the above, we begin by assuming a self-similar momentum-conserving solution $R_{s}=A_{\rm P}t^{\beta_{\rm P}}$ ($\Rightarrow v_{\rm s}=\beta_{\rm P}R_{\rm s}/t$). The equation of motion is then 
\begin{equation}
\tau_{\rm in} \frac{L_{\rm AGN}}{c} t  = M_{\rm s} v_{\rm s},
\end{equation}
yielding
\begin{equation}
A_{\rm P} = \left[
\frac{(3 - \alpha) (4 - \alpha) \tau_{\rm in} L_{\rm AGN}}{8 \pi \rho_{0} R_{0}^{\alpha} c}
\right]^{\frac{1}{4-\alpha}},
\end{equation}
\begin{equation}
\beta_{\rm P}=\frac{2}{4-\alpha},
\end{equation}
and
\begin{equation}
\label{vs P cons}
v_{\rm s} = 2 \left[ 
\frac{(3-\alpha) L_{\rm AGN} \tau_{\rm in}}{8 \pi (4 - \alpha) c \rho_{0} R_{0}^{\alpha}}
\right]^{1/2} R_{\rm s}^{(\alpha - 2)/2}~~~~~(P~{\rm cons.}).
\end{equation}
While energy-conserving solutions transition to a partially radiative bubble when $t_{\rm p,cool}/t_{\rm flow} < 1$, momentum-conserving solutions enter the partially radiative bubble phase when $t_{\rm p,cool}/t_{\rm cr} > 1$, where $t_{\rm cr} = R_{\rm sw}/v_{\rm in}$. In the momentum-conserving phase, $R_{\rm sw} \approx R_{\rm s}$, so that
\begin{align}
\frac{t_{\rm p,cool}}{t_{\rm cr}} = \frac{3^{8/5}}{8 \sqrt{2}} 
& \left[
\left( \frac{\alpha-4}{\alpha-3} \right)^{3}
\frac{\pi^{11} c^{19} m_{\rm e}^{4} m_{\rm p}^{14} \mu^{6} R_{0}^{3 \alpha} \rho_{0}^{3} v_{\rm in}^{36}}{e^{16} \sigma_{\rm T}^{6} (\ln{\Lambda})^{4} L_{\rm AGN}^{13} \tau_{\rm in}^{7}}
\right]^{1/10} \\ \notag
&~~~~~~~~~~~~~~~~~~~~~~~~~~~~~~~~~~~ \times R_{\rm s}^{(16-3\alpha)/10}.
\end{align}
This ratio increases with increasing $R_{\rm s}$ for all $\alpha<2$, so there is always a transition to a partially radiative bubble at a finite radius 
\begin{align}
R_{\rm PRB}^{\rm P} = \left[ 
\frac{3^{16} \pi^{11}}{2^{35}} \left( \frac{\alpha-4}{\alpha-3} \right)^{3}
\frac{c^{19} m_{\rm e}^{4} m_{\rm p}^{14} \mu^{6} R_{0}^{3\alpha} \rho_{0}^{3} v_{\rm in}^{36}}{e^{16} \sigma_{\rm T}^{6} (\ln{\Lambda})^{4} L_{\rm AGN}^{13} \tau_{\rm in}^{7}}
\right]^{1/(3 \alpha - 16)}.
\end{align}
Figure \ref{Rcp fig} compares $R_{\rm PRB}^{\rm P}$ with $R_{\rm f}$ for representative conditions. 
The outflow experiences a momentum-conserving phase only if $R_{\rm PRB}^{\rm P}>R_{\rm f}$; this again requires a fairly high ambient density, generally in excess of the mean density swept-up by observed galaxy-scale AGN outflows.

\section{Emission from the shocked wind bubble}
\label{shocked wind emission}
In this section, we estimate the observational signatures of the shocked wind bubble in emission.

We begin with emission from thermal electrons in the shocked wind bubble. 
For these estimates, we assume the late-time, energy-conserving limit.
The shocked wind density is then given by equation (\ref{shocked wind density}) and the electron temperature by equation (\ref{Te eq}). 
We neglect the effects of mixing and further approximate the bubble volume as $V_{\rm s} = 4 \pi R_{\rm s}^{3}/3$.  
Then, the free free luminosity
\begin{align}
L_{\rm ff} & \approx \epsilon_{\rm ff} V_{\rm s} \\ \notag 
& \approx 4.6\times10^{39}{\rm~erg~s^{-1}} \left( \frac{L_{\rm AGN}}{\rm 10^{46}~erg~s^{-1}} \right)^{2} \\ \notag 
& \times \left( \frac{v_{\rm in}}{\rm 30,000~km~s^{-1}} \right)^{-9/5} \left( \frac{R_{\rm s}}{\rm 1~kpc} \right)^{-1} \\ \notag
& \times \left( \frac{v_{\rm s}}{\rm 1,000~km~s^{-1}} \right)^{-11/5} 
\left( \frac{\ln{\Lambda}}{40} \right)^{1/5} 
C 
\bar{g}_{\rm B}
\tau_{\rm in}^{11/5},
\end{align}
with the spectrum peaking around $k T_{\rm e} \approx 260~{\rm keV}~(T_{\rm e}/3\times10^{9}~{\rm K})$. 

Similarly, we evaluate the (non-relativistic) inverse Compton luminosity, 
\begin{align}
L_{\rm IC} & \approx 4 \pi \int_{0}^{R_{\rm s}} dR R^{2} \left( \frac{4 k T_{\rm e}^{\rm eq}}{m_{\rm e} c^{2}} \right) c \sigma_{\rm T} n_{e} U_{\rm ph} \\ \notag
& \approx 9.1\times10^{41}~{\rm erg~s^{-1}} \left( \frac{L_{\rm AGN}}{\rm 10^{46}~erg~s^{-1}} \right)^{2} \\ \notag
& \times \left( \frac{v_{\rm in}}{\rm 30,000~km~s^{-1}} \right)^{-3/5} \left( \frac{R_{\rm s}}{\rm 1~kpc} \right)^{-1} \\ \notag
& \times \left( \frac{v_{\rm s}}{\rm 1,000~km~s^{-1}} \right)^{-7/5} \left( \frac{\ln{\Lambda}}{40} \right)^{2/5} \tau_{\rm in}^{7/5}.
\end{align}
The inverse Compton component should manifest itself at the much lower energies characteristic of the AGN radiation field, $\sim k T_{\rm X} \sim$ keV \citep[][]{2010MNRAS.402.1516K}. 

The synchrotron emission is sensitive to the unknown magnetic field distribution and related to the inverse Compton power via equation (\ref{P synch}). Since the thermal electrons do not have ultra-relativistic velocities, the characteristic synchrotron frequency
\begin{equation}
v_{\rm c} \sim \frac{3}{4 \pi} \frac{\gamma^{2} e B}{m_{\rm e} c} \approx 4~{\rm Hz}~\gamma^{2} \left( \frac{B}{\rm 1~\mu G} \right)
\end{equation}
is so low that the interstellar medium of the Galaxy is opaque to this emission \citep[][]{2010MNRAS.406..863L}.

Particles can also be accelerated to relativistic velocities at the forward and reverse shocks. 
These can then produce non-thermal emission including radio synchrotron, and $\gamma$-rays from inverse Compton scattering and hadronic processes. 
Although the details are uncertain, a general estimate of the non-thermal emission from the AGN shocked wind bubble can be obtained by viewing it as a supernova remnant analog and exploiting our knowledge of non-thermal emission from star-forming galaxies. 
The gigahertz radio continuum of star-forming galaxies is in fact believed to originate from cosmic ray electrons, and $\gamma-$ray emission from cosmic ray protons has also recently been detected in nearby starbursts \citep[][]{2011ApJ...734..107L}. 

As before, let $L_{\rm in} = \dot{M}_{\rm in} v_{\rm in}^{2} / 2$ be the mechanical luminosity of the AGN wind and denote by $\dot{E}_{\rm k}^{\rm SNR} \equiv f_{\rm SN} f_{\rm SN, k} L_{\star}$ the analogous quantity for SNRs, integrated over the galaxy. Here, $L_{\star}$ is the stellar bolometric luminosity, $f_{\rm SN}$ is the fraction going into supernova explosions, and $f_{\rm SN,k}$ is the fraction of the supernova energy going into kinetic power (rather than being radiated away). 
Then:
\begin{align}
\frac{L_{\rm in}}{\dot{E}_{\rm k}^{\rm SNR}} & = \left( \frac{\tau_{\rm in} v_{\rm in}}{2 f_{\rm SN} f_{\rm SN,k} c} \right) \frac{L_{\rm AGN}}{L_{\star}} \\ \notag
& = 5 \tau_{\rm in} \left( \frac{v_{\rm in}}{0.1c} \right)  \left( \frac{f_{\rm SN}}{0.01} \right)^{-1} \left( \frac{f_{\rm SN,k}}{1} \right)^{-1} \frac{L_{\rm AGN}}{L_{\star}},
\end{align}
where the fiducial value for $f_{\rm SN}$ is based on the results of \cite{1982ApJ...263..723A}. 
Assuming a comparable efficiency for accelerating particles in SNR and AGN wind shocks, this shows that the non-thermal emission from the AGN wind bubble can in principle significantly exceed that powered by star formation. In a ULIRG like Mrk 231 with a powerful quasar at the center, $L_{\rm AGN} \sim L_{\star}$ \citep[e.g.,][]{2009ApJS..182..628V}. 
We caution, however, that this ratio should be taken with a grain of salt since the conditions near the shock acceleration sites in SNRs and in the shocked wind bubble (including the photon and magnetic energy densities, and the time scales for advection and adiabatic losses) need not be the same. 

Of particular interest is the non-thermal synchrotron contribution, which may have been observed in some systems. 
An upper bound can derived under the assumption that the system acts as a calorimeter, i.e. that the cosmic ray electrons radiate all their energy in synchrotron on a time scale short relative to all other radiative and dynamical time scales.  
Assuming further that the particles are shock-accelerated with a spectrum $n(\gamma) \propto \gamma^{-p}$ with $p=2$ \citep[][]{1987PhR...154....1B}, the synchrotron emission has equal power per frequency decade,
\begin{align}
\label{nonthermal synch}
\left. \nu L_{\rm \nu} \right|_{\rm synch}^{\rm max} & \approx \frac{\xi L_{\rm in}}{\ln{\gamma_{\rm max}}} = \frac{\xi \tau_{\rm in}}{2 \ln{\gamma_{\rm max}}} \left( \frac{v_{\rm in}}{c} \right) L_{\rm AGN} \\ \notag
& \approx 3.6\times 10^{41}~{\rm erg~s^{-1}}~\tau_{\rm in} \left( \frac{\xi}{0.01} \right) \left( \frac{\ln{\gamma_{\rm max}}}{\ln{10^{6}}} \right)^{-1} \\ \notag
&~~~~~~~~~~~~~~~~~~~~~~~~~ \times \left( \frac{v_{\rm in}}{0.1c} \right) \left( \frac{L_{\rm AGN}}{\rm 10^{46}~erg~s^{-1}} \right).
\end{align}
The efficiency factor $\xi \approx 1$\% is constrained for SNR shocks based on modeling the far infrared-radio correlation \citep[e.g.,][]{2010ApJ...717....1L}. 
The synchrotron luminosity estimated in equation (\ref{nonthermal synch}) can only be realized if $U_{\rm B}>U_{\rm ph}$, which may not be realized given the large photon densities near AGN (see eq. (\ref{synch to Comp ratio})). 

\section{Modeling the effects of fast nuclear winds in simulations}
\label{numerical simulations}
It is at present extremely challenging to simultaneously resolve the launching of AGN winds and their effects on galaxy scales in numerical simulations. 
For example, the accretion disk wind model of \cite{1995ApJ...451..498M} predicts that winds are launched at $\sim 0.01$ pc from the black hole. 
However, one can model their effects in galaxy simulations by imparting kinetic kicks to gas elements, and attempt to resolve the resulting wind bubble \citep[e.g.,][]{2011MNRAS.tmp.2150D}. 
In this section, we use our analytic model to outline the basic requirements for numerical simulations to faithfully capture the impact of fast nuclear winds on galaxy scales. It is not our intention to provide a complete algorithmic description, as this would depend on the details of the numerical scheme used, but rather to summarize the criteria that realistic implementations must achieve.

As discussed in \S \ref{shocked wind cooling}, it is the cooling of the shocked wind, rather than the shocked ambient medium, that determines whether the outflow conserves energy or momentum. 
Thus, a realistic simulation should at minimum resolve the wind shock. 
The mass of the shocked wind is
\begin{align}
\label{Msw}
M_{\rm sw} & \sim \dot{M}_{\rm in} \frac{R_{\rm s}}{v_{\rm s}}  \\ \notag 
& \sim 2.2\times 10^{6}~{\rm M_{\odot}}~f_{\rm Edd} \left( \frac{\eta}{0.1} \right)^{-1}
\left( \frac{M_{\rm BH}}{\rm 10^{8}~M_{\odot}} \right) \\ \notag 
&~~~~~ \times \left( \frac{R_{\rm s}}{\rm 1~kpc} \right) \left( \frac{v_{\rm s}}{\rm 1,000~km~s^{-1}} \right)^{-1}.
\end{align} 
In the energy-conserving phase, the radius of the wind shock is
\begin{align}
R_{\rm sw} & \sim R_{\rm s} \left( \frac{v_{\rm s}}{v_{\rm in}} \right)^{1/2} \\ \notag 
& \sim 180~{\rm pc} \left( \frac{R_{\rm s}}{\rm 1~kpc} \right) \left( \frac{v_{\rm s}}{\rm 1,000~km~s^{-1}} \right)^{1/2} \\ \notag
&~~~~~  \times \left( \frac{v_{\rm in}}{\rm 30,000~km~s^{-1}} \right)^{-1/2},
\end{align}
whereas $R_{\rm sw} \sim R_{\rm s}$ when it is momentum-conserving. 
However, a significantly better resolution than suggested by the above estimates is needed in general to accurately capture the cooling time of the shocked wind. 
In particular, many numerical codes broaden shocks over several resolution elements, which allows gas to artificially cool while it is being shocked \citep[][]{2000MNRAS.319..721H, 2011MNRAS.415.3706C}. 
As a result, the simulated gas may never reach the physical post-shock temperature 
\begin{equation}
\label{Tsw}
T_{\rm sw} \sim 10^{10}~{\rm K}\left( v_{\rm in}/30,000{\rm~km~s^{-1}} \right)^{2}.
\end{equation} 
 
Thus, numerical simulations should:
\begin{itemize}
\item Ensure that a wind shock is present, with radius, shocked wind mass and temperature consistent with equations (\ref{Msw}-\ref{Tsw}) in the energy-conserving phase.
\item Implement a cooling function extending to $T>10^{10}$ K, including Compton scattering. Two-temperature plasma effects in the shocked wind can be modeled following the results of \S \ref{shocked wind cooling}. The cooling time of the shocked wind should be checked against the analytic expectations to diagnose potential artificial losses due to numerical effects.
\item The time integration scheme must ensure accurate energy conservation at the numerical level. For instance, in the absence of a proper limiter neighboring smooth particle hydrodynamics particles that are initially at rest may not react adequately to sudden changes, which can lead to effects including inter-particle crossing and violation of energy conservation \citep{2009ApJ...697L..99S, 2012MNRAS.419..465D}.
\end{itemize}
If it is not possible to correctly resolve the dynamics of the wind bubble, approximations that enforce physical energy conservation may be considered. For instance, some implementations of supernova feedback suppress cooling in order to avoid efficient energy radiation when the ISM is not properly resolved (for a review, see Kay et al. 2002; Dalla Vecchia \& Schaye 2012 discuss alternatives\nocite{2002MNRAS.330..113K, 2012arXiv1203.5667D}), and other techniques have also been developed in the context of AGN feedback \citep[e.g.,][]{2007MNRAS.380..877S, 2009MNRAS.398...53B}. 

\bibliography{references} 
 
\end{document}